\documentclass[twoside,british]{elsarticle}
\usepackage[T1]{fontenc}
\usepackage[latin9]{inputenc}
\usepackage{color}
\usepackage{verbatim}
\usepackage{url}
\usepackage{amssymb}
\usepackage{wasysym}
\usepackage{graphicx}

\makeatletter

\providecommand{\tabularnewline}{\\}

\journal{Example: NIM A}
\usepackage{lineno}

\makeatother

\usepackage{babel}
\begin{document}

\title{A Helium Gas-Scintillator Active Target for Photoreaction Measurements}

\begin{frontmatter}{}

\author[gla]{R. Al Jebali}

\author[gla]{J.R.M. Annand\corref{cauth}}

\ead{john.annand@glasgow.ac.uk}

\author[lu]{J.-O. Adler}

\author[isp]{I. Akkurt}

\author[gla]{E. Buchanan}

\author[lu1]{J. Brudvik}

\author[lu]{K. Fissum}

\author[gla]{S. Gardner}

\author[gla]{D.J. Hamilton}

\author[lu1]{K. Hansen}

\author[lu1]{L. Isaksson}

\author[gla]{K. Livingston}

\author[lu1]{M. Lundin}

\author[gla]{J.C. McGeorge}

\author[gla]{I.J.D. MacGregor}

\author[gla]{R. MacRae}

\author[mz]{D.G. Middleton}

\author[gla]{A.J.H. Reiter\fnref{gsi}}

\author[gla]{G. Rosner\fnref{fair}}

\author[lu1]{B. Schröder}

\author[gla,lu1]{J. Sjögren}

\author[gla]{D. Sokhan}

\author[gla]{B. Strandberg}

\fntext[gsi]{Present Address: Gesellschaft für Schwerionenforschung mbH, Planckstrasse
1, 64291 Darmstadt, Germany}

\fntext[fair]{Present Address: FAIR, Facility for Antiproton and Ion Research in
Europe GmbH, Planckstr. 1, D-64291, Germany}

\cortext[cauth]{Corresponding author}

\address[gla]{School of Physics \& Astronomy, University of Glasgow, G12 8QQ, Scotland,
UK}

\address[lu]{Department of Physics, University of Lund, Sölvegatan 14, SE-223
62, Lund, Sweden}

\address[lu1]{MAX IV Laboratory, PO Box 118, SE-221 00, Lund, Sweden}

\address[isp]{Süleyman Demirel University, Fen-Edebiyat Faculty, 32 260 Isparta,
Turkey}

\address[mz]{Kepler Centre for Astro and Particle Physics, Physikalisches Institut,
Universität Tübingen, D-72076 Tübingen, Germany}
\begin{abstract}
A multi-cell He gas-scintillator active target, designed for the measurement
of photoreaction cross sections, is described. The target has four
main chambers, giving an overall thickness of 0.103~$\mathrm{g/cm^{2}}$
at an operating pressure of 2~MPa. Scintillations are read out by
photomultiplier tubes and the addition of small amounts of $\mathrm{N}_{2}$
to the He, to shift the scintillation emission from UV to visible,
is discussed. First results of measurements at the MAX IV Laboratory
tagged-photon facility show that the target has good timing resolution
and can cope well with a high-flux photon beam. The determination
of reaction cross sections from target yields relies on a Monte Carlo
simulation, which considers scintillation light transport, photodisintegration
processes in $^{4}\mathrm{He}$, background photon interactions in
target windows and interactions of the reaction-product particles
in the gas and target container. The predictions of this simulation
are compared to the measured target response.
\end{abstract}

\end{frontmatter}{}

\begin{comment}
\linenumbers
\end{comment}

\section{\label{sec:Introduction}Introduction}

\textcolor{black}{The group-eight elements scintillate in the gas
and liquid phases \citep{Birks}, producing a signal that has a linear
dependence on energy deposited. Unlike many scintillators, there is
no strong velocity dependence of the signal so that relatively low
velocity heavy ions can be detected.} Since the 1950's \citep{Eggler}
a variety of inert-gas scintillators have been developed for the detection
of charged ions, but much of the effort has concentrated on Xe or
Ar, which give good stopping power and high gain if the gas is also
used as the ionisation medium for a proportional counter. More recently
liquid Xe and Ar have been used for high-energy, hadron calorimetry
\citep{LHC}.

Detectors using He scintillator tend to be more specialised, often
driven by the desire to investigate the properties of $^{3}\mathrm{He}$
or $^{4}\mathrm{He}$ nuclei. High-pressure $^{4}\mathrm{He}$ gas
cells \citep{MorganWalter} have been used for fast neutron polarimetry
\citep{RBG,Drigo} as the analysing power for $n-{}^{4}\mathrm{He}$
scattering is large and well known. Liquid helium scintillators, also
with a long pedigree \citep{Thorndike}, have similarly been used
for neutron polarimetry \citep{Kazuo} and more recently to measure
beta decay of magnetically trapped ultra-cold neutrons \citep{McKinsey}.

In this article we report on the development of a $^{4}\mathrm{He}$
gas-scintillator active target (AT), where the target material is
also the detection medium for the charged products of nuclear reaction
processes. The objective is to measure the total and partial $\mathrm{^{4}He}$
photodisintegration cross sections at photon energies from breakup
threshold, potentially up to pion production threshold. These observables
are sensitive to the structure of the $\mathrm{^{4}He}$ nucleus and
are important to the development of \emph{ab initio} methods to calculate
the $\mathrm{^{4}He}$ wave function. The existing data set is surprisingly
patchy \citep{Quaglioni} and has offered often contradictory evidence
to these fundamental theoretical efforts.

\textcolor{black}{With a conventional separate target and detector
arrangement, it is difficult to reach the near-threshold region since
low energy charged ions are easily stopped. Thus an AT, where the
ions do not have to pass through any inert material before detection,
is highly advantageous. A different AT technique where the He gas
(with 25\% methane admixture) is the ionisation medium for a time
projection chamber has been used \citep{Shima} to explore similar
physics issues. The present target does not employ tracking elements
for event reconstruction, so that large admixtures of other gases
are undesirable if the He cross section is to be measured accurately.}

The following sections describe the construction of the AT, the obtained
scintillation signal, Monte Carlo (MC) simulations and the first measurements
of $\gamma+^{4}\mathrm{He}$ reactions using the tagged photon beam
at the MAX IV Laboratory.

\section{\label{sec:Bench Tests}The Gas Scintillator Active Target}

\begin{figure}
\includegraphics[clip,width=1\columnwidth]{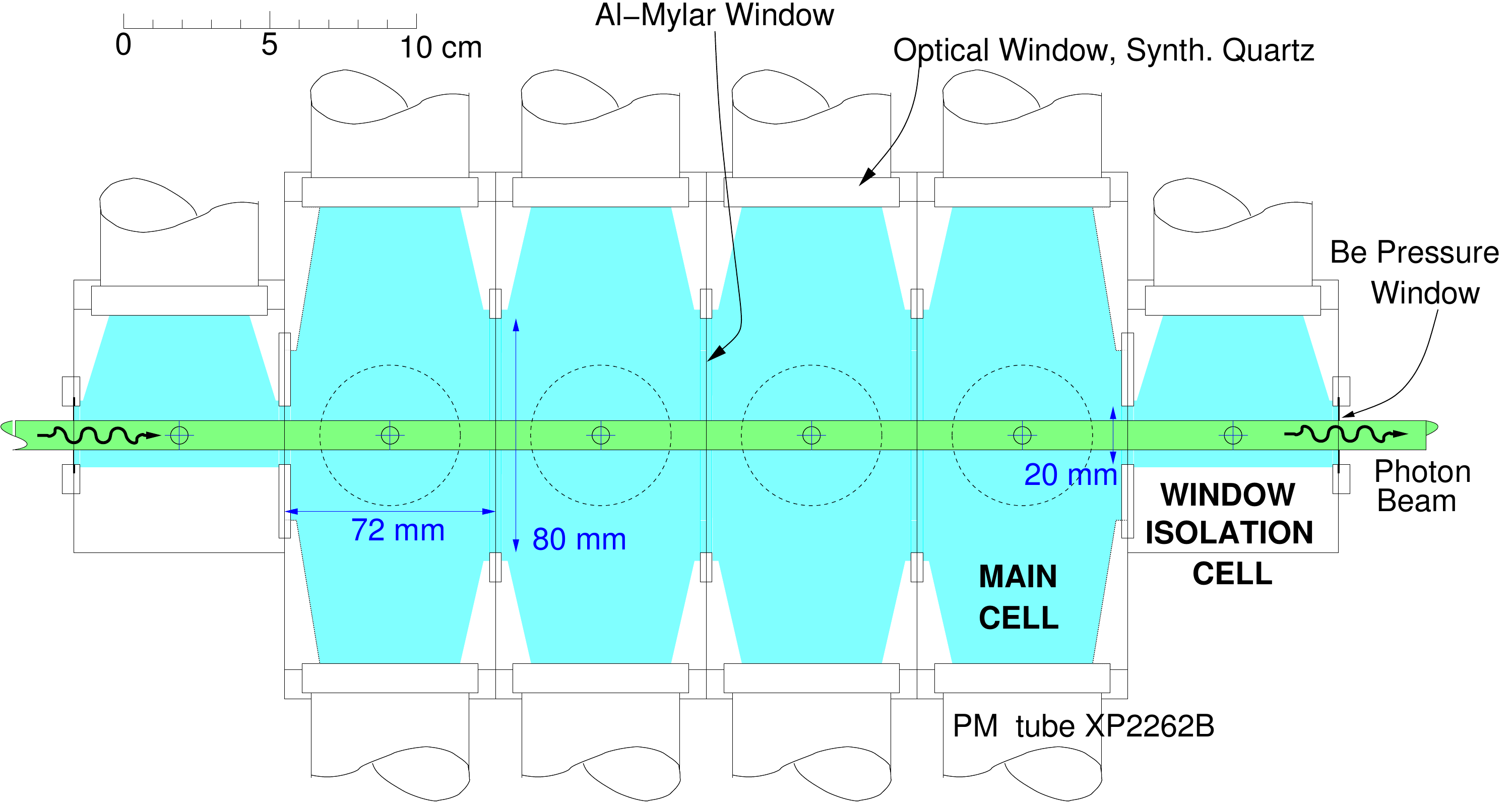}

\protect\caption{\label{target_plan}Plan view of the active target.}
\end{figure}

The active target is shown schematically in Fig.\ref{target_plan}
and is described in detail in Ref.\citep{Jebali:2013}. It consists
of 
\begin{itemize}
\item 4 \textit{Main} \emph{Cells}. Each cell is read out by 4 photomultiplier
tubes (PMT), viewing the gas chamber through 10~mm thick synthetic
quartz windows. A pressure-tight seal is made between the window and
the body of the target using indium gaskets. The cell length is 72~mm,
giving a thickness of 0.0257~$\mathrm{g/cm^{2}}$ per cell at 2~MPa
pressure. The joint between cells is sealed by a ``V-ridge'' on
a Cu gasket and the pressure between cells is equalised via an internal
passage. The cells are isolated optically by 5~$\mu$m thick aluminised
Mylar windows.
\item 2 \textit{Window Isolation} \emph{Cells}. These cells, attached at
either end of the target, have a single PMT attached to the target
body in the same manner as the main-cell PMT. They isolate the main
cells from particles produced in the outer pressure-containment windows,
which are 0.5~mm thick Be. The main utility of the isolation cells
is to remove the Be window from close proximity to the main cells.
Signals from the isolation cells can be used to veto events from beam
interaction with the windows.
\end{itemize}
The cells of the target were machined from solid Al alloy, cleaned
and coated on the internal surfaces with $\mathrm{TiO_{2}}$ reflector.
The AT was then assembled and pressure tested with 2~MPa He. When
leak tight the PMTs were added, coupled to the windows via optical
grease. 

\textcolor{black}{}

\subsection{\label{sub:Gas-Handling}Gas Handling}

Prior to filling with He, the target and all the gas handling apparatus
were evacuated. Any trace non-helium gas (for scintillation wavelength
shifting) was then introduced and finally the He was added. During
measurements with beam the target pressure and temperature were monitored
continuously.

\begin{comment}
\begin{figure}
\begin{center}\includegraphics[width=1\columnwidth,bb = 0 0 200 100, draft, type=eps]{Plots/ATGasSystem.jpeg}\end{center}

\protect\caption{\label{fig:gas-handle}Schematic view of gas handling system. The
nitrogen and xenon loops allow trace amounts of gas to be added to
the target cell via a small ancillary chamber (Circuit III).}
\end{figure}
\end{comment}

\subsection{\label{sub:helium-scintillation}Investigation of Helium Scintillation}

The scintillation emission spectrum of helium, in common with all
noble gases, is a complex system of lines, bands and continua. The
spectrum extends from the Near Infra-red (NIR) into the Vacuum Ultra
Violet (VUV, $\lambda<100$~nm) \citep{HeRefractiveIndex} and depends
on pressure, modes of excitation and the ionisation density of the
incident charged particle. At low pressure ($\leq0.1$ kPa) the scintillation
is produced by atomic processes, while at higher pressure, collisions
between excited or ionised He with neighbouring unexcited atoms, can
produce He eximers. Part of the eximer's energy is released in the
form of scintillation, predominantly in the VUV. The intensity of
the scintillation is expected \citep{Birks} to be of the order $10^{3}$
photons per MeV deposited. VUV is difficult to detect with good efficiency
and it is common to add an impurity to the He. Collisions and other
processes, transfer energy from He eximers and/or excited atoms to
the impurity molecules, which then emit at their longer characteristic
wavelengths. $\mathrm{N_{2}}$ and Xe are commonly used impurities.
Although this results in an overall reduction in scintillation efficiency
\citep{Birks}, this is more than compensated by the improved reflection,
transmission and PMT quantum efficiency for visible photons. Furthermore,
standard glass-window PMTs can be used, in this case XP2262B, which
are cheaper and less susceptible to helium ingress than quartz.

Initial investigations of helium scintillation properties were made
with a small test cell, which housed an open $^{241}\mathrm{Am}$
$\alpha$ source (energy $\sim5.5$~MeV), where the emissions were
viewed by a quartz-window PMT type XP2020Q. The small amplitude of
the pulses confirmed that transport and collection of the UV scintillation
is inefficient, even with quartz windows on the target and a quartz-window
PMT.

A wavelength-shifting ``paint'' type EJ-298 \citep{EJ-298}, which
consists of a polyvinyl toluene binder and $\mathrm{C_{2}H_{4}(CH_{3})_{2}}$
fluorescent dopant dissolved in xylene, was then used to coat the
quartz window and the internal surfaces of the cell, which were previously
coated with $\mathrm{TiO_{2}}$ reflector. The dopant gives peak emissions
at $\sim425$~nm and is commonly used to shift the primary UV scintillation
in plastic scintillator. The paint did boost the He scintillation
yield, but also produced a signal itself, as was observed when the
cell was evacuated, allowing the $\alpha$ particles to strike the
cell walls. The paint signal was considerably stronger than that from
the shifted He scintillation.

\begin{figure}
\begin{center}\includegraphics[width=1\columnwidth]{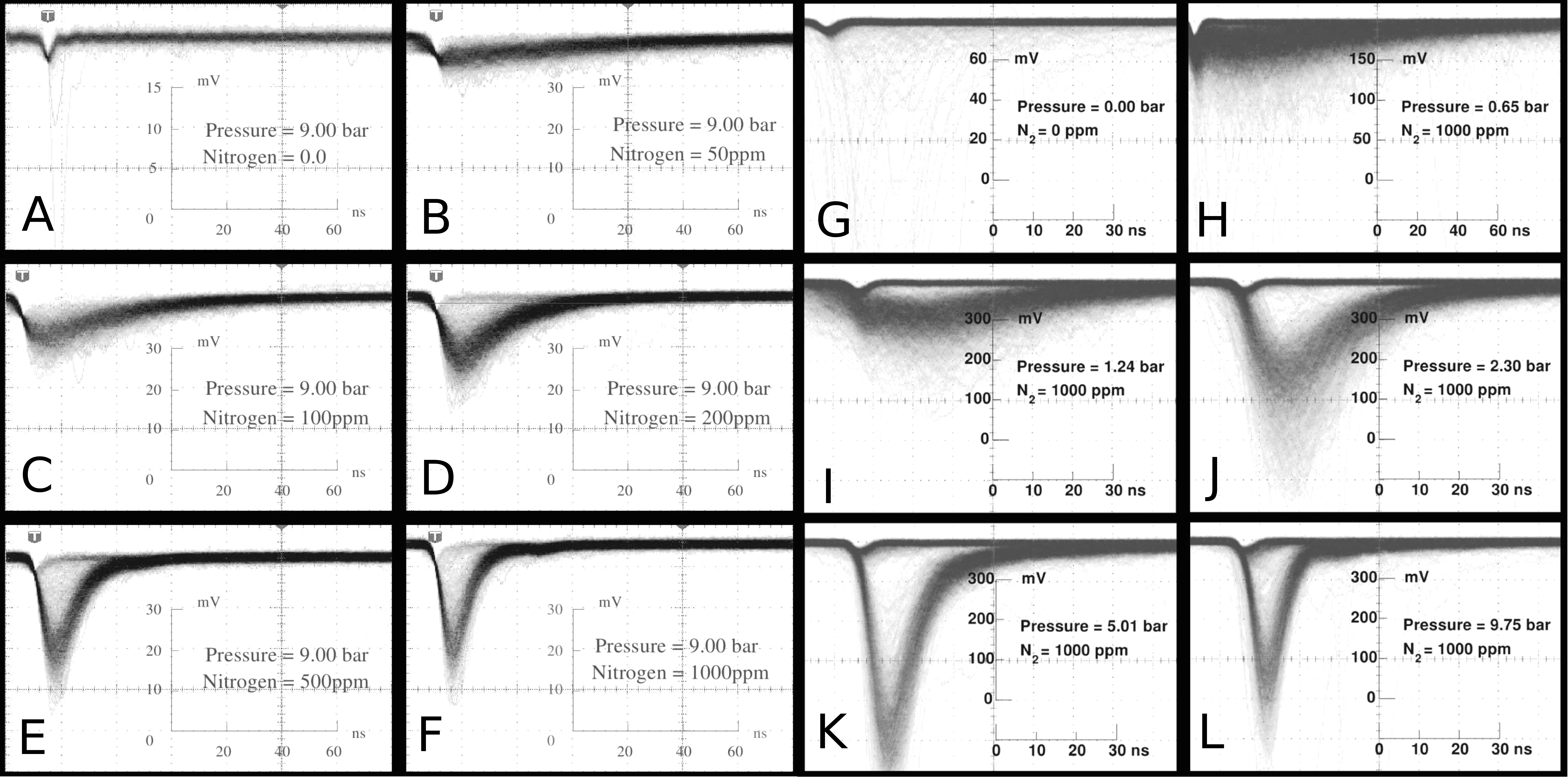}\end{center}

\protect\caption{\label{N2conc}He gas scintillation pulses for incident $\sim5.5$~MeV
$\alpha$ particles. A-F: constant pressure of 0.9~MPa (9~bar) with
$\mathrm{N}_{2}$ concentrations 0, 50, 100, 200, 500, 1000~ppm respectively.
G: Empty Cell. H-L: fixed $\mathrm{N_{2}}$ concentration at 1000~ppm
and pressures 0.065, 0.124, 0.230, 0.501, 0.975~MPa respectively.}
\end{figure}

\begin{comment}
\begin{figure}
\begin{center}\includegraphics[width=1\columnwidth,bb = 0 0 200 100, draft, type=eps]{Plots/PulseXeConc.jpeg}\end{center}

\protect\caption{\label{fig:XeConc}Pulse forms for $\sim5.5$~MeV $\alpha$ particles,
for different concentrations of Xe and Xe + $\mathrm{N_{2}}$.}
\end{figure}
\end{comment}

\begin{comment}
\begin{figure}
\begin{center}\includegraphics[width=1\columnwidth,bb = 0 0 200 100, draft, type=eps]{Plots/PulseHePressure.jpeg}\end{center}

\protect\caption{\label{HePressure}Pulse forms for $\sim5.5$~MeV $\alpha$ particles
at different values of pressure. $\mathrm{N_{2}}$ concentration was
fixed at 1000~ppm.}
\end{figure}
\end{comment}

Alternatively a trace of impurity gas was added to the He. Tests were
performed using $\mathrm{N_{2}}$, Xe and a mixture of $\mathrm{N_{2}}$
and Xe, to shift the primary scintillation to the $\sim420$ nm range,
which is optimum for the bialkalai-cathode XP2262B PMT. Fig. \ref{N2conc}A-F
show oscilloscope traces for the $\alpha$ response as a function
of $\mathrm{N_{2}}$ concentration at a pressure of $\sim0.9$~MPa.
It can be seen that the pulse decay time decreases as the $\mathrm{N_{2}}$
concentration is increased up to 1000~ppm. With Xe admixtures at
much higher concentration (2 - 20\%) the $\alpha-\mathrm{particle}$
peak is less well defined. Adding 500~ppm $\mathrm{N_{2}}$ along
with the Xe does not improve the performance and Xe admixtures were
not investigated further.

The response also depends on pressure and Fig. \ref{N2conc}H-L show
that pulse rise and fall times decrease as the pressure was raised
to $\sim1$~MPa (9.75~bar), although the amplitude dropped between
0.5 and 1~MPa. There was a further, relatively small reduction in
amplitude when the pressure was raised to 2~MPa, but the pulse time
dependence did not change significantly compared to the 1~MPa case.

Although the gamma-ray detection efficiency is very low, the AT sitting
directly in an intense bremsstrahlung photon beam will generate relatively
high counting rates, so that a short pulse length is desirable. Furthermore
since the goal of the project is precise measurement of $^{4}\mathrm{He}$
photoreaction cross sections, any wavelength-shifting impurity concentration
should be kept small. At 1000~ppm $\mathrm{N_{2}}$ concentration
the effect on a total cross section measurement will generally be
very small (Sec. \ref{sub:BeWindowSim}), although at energies very
close to $^{4}\mathrm{He}$ photodisintegration threshold ($E_{\gamma}\sim20$~MeV)
the $\mathrm{N_{2}}$ background may be larger. With this $\mathrm{N_{2}}$
concentration and 2~MPa pressure, a well-defined, sharp signal, with
a rise time of $\sim5$~ns and fall time of 10 ns, was observed for
$\sim5.5$~MeV $\alpha$ particles. These operating conditions were
used for the tagged photon experiment at MAX IV Laboratory (Sec.\ref{sec:MAX measurements}).

\section{\label{sec:Monte-Carlo-Models}Monte Carlo Models}

The response of the AT has been simulated using a MC model based on
Geant-4 \citep{G4-Phys}. This was performed in two stages. The first
(Sec. \ref{sub:Optical-Photon-Transport}) calculates the transport
of scintillation photons from the point of their creation up to the
point where they produce photoelectrons in the cathode of a PMT. From
this the position dependence of the amplitude of the PMT signal is
obtained. The second (Sec. \ref{sub:Photo-Reaction-Modeling}) calculates
the energy deposited by photodisintegration products for two, three
and four-body breakup of $\mathrm{^{4}He}$. From that, and the results
of the optical transport simulation, the amplitudes of the AT signals
at the PMT photocathodes are calculated. This procedure also provides
estimates of the effect of materials other than He on the AT signal.

\subsection{\label{sub:Geometry}AT Geometry Specification}

\begin{figure}
\begin{center}\includegraphics[width=1\columnwidth]{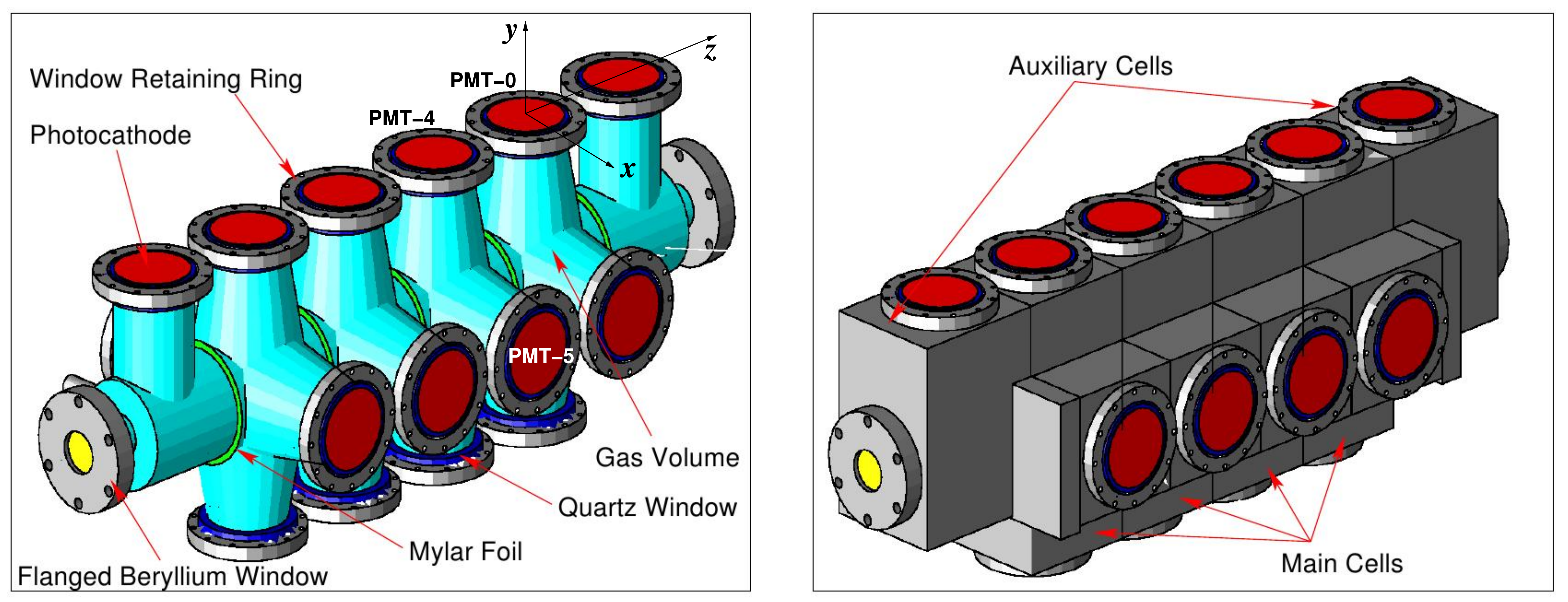}\end{center}

\protect\caption{\label{fig:Active-Target-geometry}Active Target geometry rendered
by the MC model. The left panel shows the internal volume filled with
He gas (cyan), while the right panel shows the outer Al body of the
AT. }
\end{figure}

Originally the Geant-4 geometry was specified in ``Geometry Dependent
Markup Language'', derived from the CAD files used to machine the
AT components. This method was discarded as the Geant-4 tracking algorithms
did not recognise AT boundaries correctly, possibly due to perceived
``overlapping volumes''.

An alternative procedure was developed using combinations of simple
shapes (Fig. \ref{fig:Active-Target-geometry}). This models the AT
geometry accurately, considering: main cells, window-isolation cells,
internal reflective paint (EJ-510), internal Al Mylar foils, quartz
windows (HOQ-310), PMT (only the glass window and photo-cathode are
modelled), Al window retaining rings and flanged Be windows. This
implementation of the geometry behaved correctly with respect to tracking
behaviour at the boundaries between volumes.

\subsection{\label{sub:Optical-Photon-Transport}Optical Photon Transport}

\begin{figure}
\begin{center} \includegraphics[width=0.9\columnwidth]{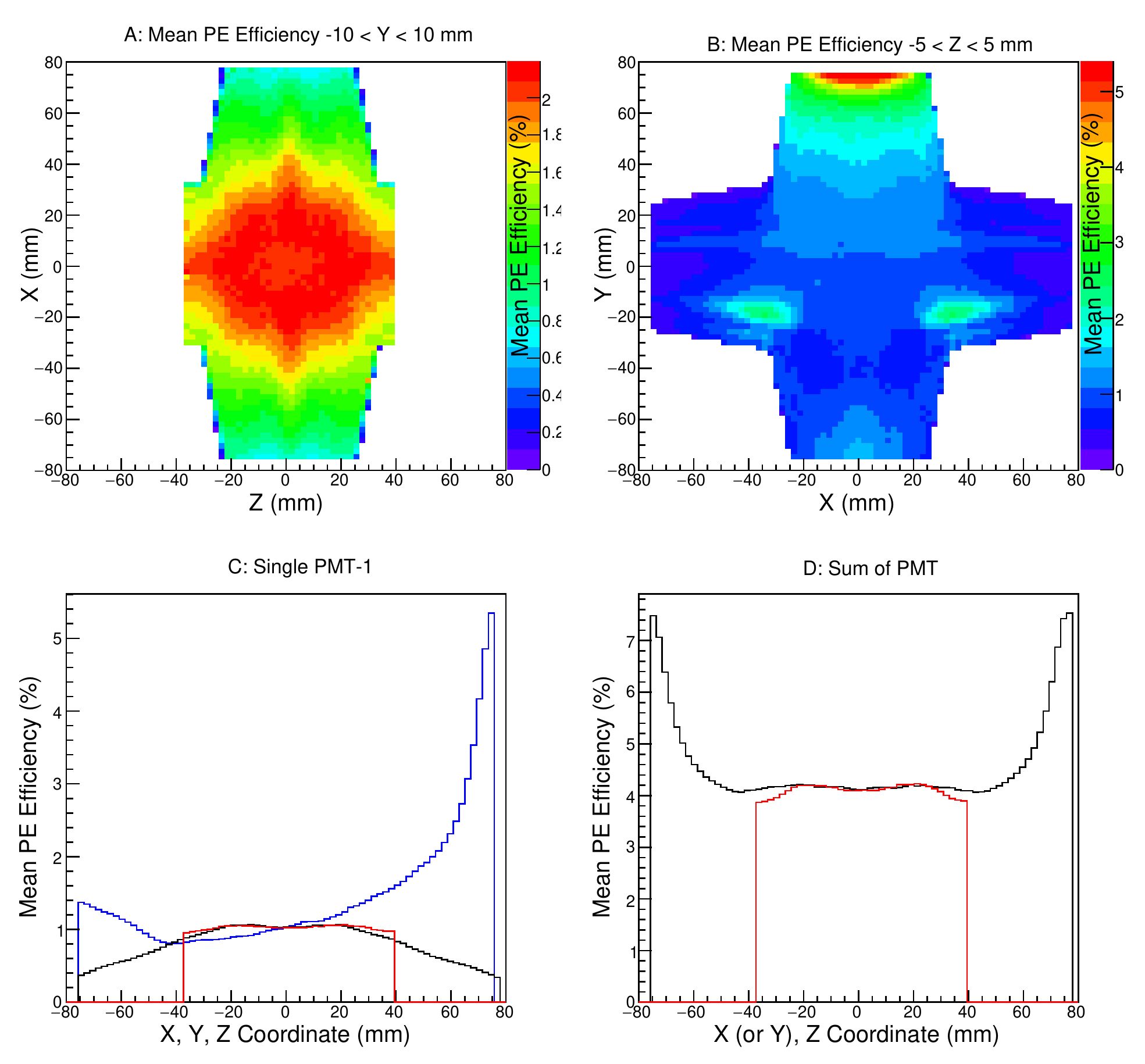}\end{center}

\protect\caption{\label{fig:Light-Collection-efficiency}PE-efficiency as a function
of position in the AT. A,B: 2D projections of the 3D PE-efficiency
map. A: x-z projection for $-10<y<+10$~mm, B: x-y projection for
$-5<z<+5$~mm. Both A and B refer to the single-PMT PE-efficiency.
C,D: 1D projections of the 3D efficiency map along the x, y and z
axes. Points within 10~mm radius of the axes are included. C: Single
PMT, Black x, Blue y, Red z. D: Sum of the 4 PMTs, Black x or y, Red
z.}
\end{figure}

For simulations involving the transport and tracking of scintillation
photons, optical properties and boundary characteristics were defined
as described in the following.
\begin{itemize}
\item The reflectivity of the paint EJ-510 \citep{EJ-298} was input as
a function of incident wavelength. Diffuse reflection from the matt
surface of the paint was modelled using the ``unified'' model of
Geant-4 \citep{G4-Phys}.
\item The absorption length, reflectivity and refractive index were input
as functions of photon wavelength for the synthetic quartz windows
(HOQ-310) \citep{hoq310} of the AT and the glass windows of the PMTs
\citep{Photonis} .
\item The He gas refractive index was taken from the calculations of Ref.\citep{HeRefractiveIndex}.
This is very close to 1 and had negligible effect on the calculation.
\item The transmission and reflectivity of the Al-Mylar foils were measured
using a UV-Visible spectrophotometer. The reflectivity is $\sim95\%$
and transmission consistent with zero at wavelengths in the region
of the $\mathrm{N_{2}}$ emission spectrum.
\item The quantum efficiency of the XP2262 photocathode \citep{Photonis}
was input as a function of incident wavelength.
\end{itemize}
The position dependence of the scintillation signal from the AT was
simulated by generating optical photons, at wavelengths sampled according
to the $\mathrm{N_{2}}$ emission spectrum \citep{N-spectrum}, at
given positions within the AT. Optical photons were tracked and the
number of photoelectrons generated in each PMT counted. The starting
position was stepped on an $x,y,z$ grid (Fig. \ref{fig:Active-Target-geometry}:
$x$ perpendicular PMT-0 axis and the AT axis, y along PMT-0 axis,
$z$ along the AT axis) and the variation in photo-electron generation
efficiency (PE-efficiency) along these directions is displayed in
Fig.\ref{fig:Light-Collection-efficiency}. PE-efficiency is the number
of photo-electrons produced in a single PMT cathode, expressed as
a percentage of the number of scintillation photons started from a
given grid point. Thus the quantum efficiency of the cathode, which
peaks at $\sim27\%$ for wavelengths in the vicinity of 400~nm, is
included. A total of $8\times10^{9}$ photons were generated, at positions
sampled throughout the volume occupied by the gas in an AT cell. The
volume was divided into voxels, each of size $2\times2\times1$~mm,
situated on a $75\times75\times75$ grid and, for each voxel, the
number of photo electrons generated in the cathode of each of the
4 PMTs was recorded. As would be expected, the highest summed PE-efficiency
occurs close to the PMT windows. Away from these regions the variation
in summed PE-efficiency is a smooth and relatively slowly varying
function of position.

Based on the measured PMT gain, the signal amplitude produced by $\sim5.5$~MeV
$\alpha$ particles is consistent with a position-averaged signal
at a single PMT cathode of $\sim5$ photo electrons per MeV.

\subsection{\label{sub:Photo-Reaction-Modeling}Photo-Reaction Modelling}

Samples of $^{4}\mathrm{He}$ photodisintegration events, which include
2, 3 and 4-body breakup modes, were produced in an external, ROOT-based
\citep{root} event generator. Event sampling used the incident-energy-dependent,
partial cross sections for $\gamma+^{4}\mathrm{He}\rightarrow p+^{3}\mathrm{H}$
($19.8$~MeV), $\gamma+^{4}\mathrm{He}\rightarrow n+^{3}\mathrm{He}$
($21.6$~MeV), $\gamma+^{4}\mathrm{He}\rightarrow p+n+d$ ($26.0$~MeV)
and $\gamma+^{4}\mathrm{He}\rightarrow2p+2n$ ($28.3$~MeV) as given
in Ref.\citep{Quaglioni}, which reviewed available data and provided
\emph{ab initio} calculations for the two-body breakup channels. The
quantities in parentheses are the reaction threshold energies. We
did not include the $\gamma+^{4}\mathrm{He}\rightarrow d+d$ channel
as the cross section is negligible compared to the others. Event sampling
was further weighted according to the bremsstrahlung ($\sim1/E_{\gamma}$)
distribution. The angular distributions of the final-state particles
were sampled from available kinematic phase space in the center-of-mass
system and then boosted into the laboratory frame. Sampling from a
$\sin^{2}\theta_{cm}$ distribution was also performed for the two-body
breakup channels, to test the sensitivity of the response to the input
angular distributions. Photon interaction points were chosen randomly
along the length of the AT, within a cone of half-angle 1.1~mr (defined
by the experimental photon collimator), with the angle sampled from
the bremsstrahlung angular distribution (Ref.\citep{G4-Phys}: Koch
and Motz distribution 2BS).

Photo-reaction events were then run through the Geant-4 model of the
AT. The ionisation energy loss of charged reaction products was calculated
using the Bethe-Bloch formula, or parametrised models \citep{G4-Phys}
at low energy where the Bethe-Bloch formalism breaks down. Multiple
scattering and other processes were also modelled using the Geant-4
``Low Energy Electromagnetic'' package \citep{G4-Phys}. At the
energies employed in the present investigation, hadronic interactions
of reaction products in the He gas, target windows and target walls
do not produce a large effect. They were however accounted for using
the ``High Precision neutron model'', ``Pre-compound model'' and
``Low energy parametrised model'' which are included in Geant-4
\citep{G4-Phys}.

As charged particles were tracked in discrete steps through the AT
gas, the position of each step was obtained. From this, the probability
was calculated of a scintillation photon generating a photoelectron
at the cathode of each of the four PMTs. This was performed by three-dimensional
interpolation from the grid of values obtained in the simulation of
optical photon transport (Sec. \ref{sub:Optical-Photon-Transport}).
The individual PMT signal amplitudes were derived from the energy
losses in the steps, weighted by the interpolated light collection
efficiencies to each PMT. These weighted energy losses were accumulated
along the track for each PMT, so that the position dependence of the
light collection efficiency was folded with the energy loss, and then
converted to the (nearest integer) number of photoelectrons. Random
fluctuations in this number were simulated by sampling from a Poisson
distribution, and the resultant converted back to energy. Electronic
noise was modelled by sampling from a Gaussian of width 0.15~MeV,
consistent with the observed width of QDC pedestal distributions. 

\begin{figure}
\begin{center}\includegraphics[width=1\columnwidth]{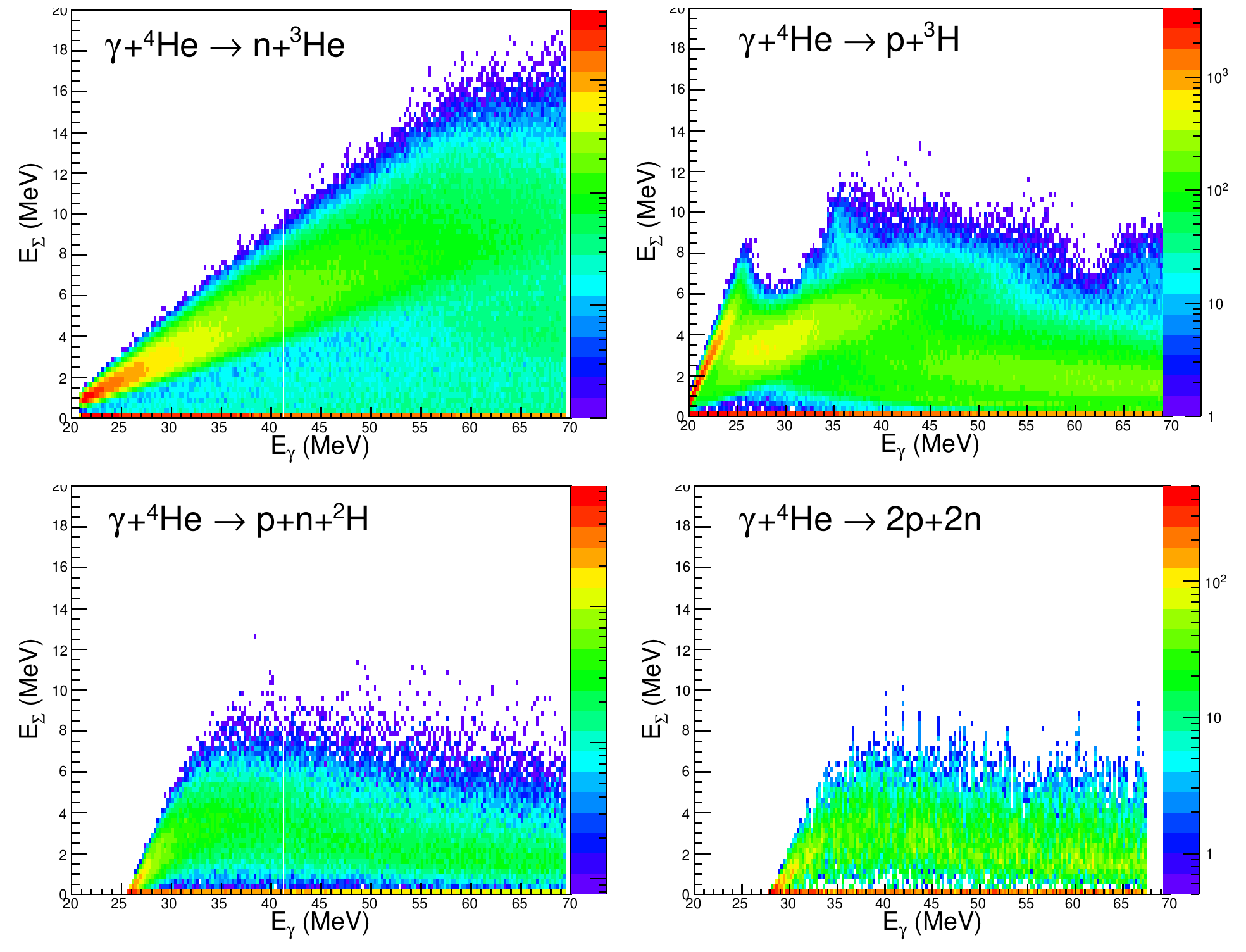}\end{center}

\protect\caption{\label{fig:energy-loss}The calculated signal amplitudes in the AT
for the four main photo-disintegration modes of $^{4}\mathrm{He}$,
shown as a function of incident photon energy $E_{\gamma}$. Pulse
height resolution effects, due to photoelectron statistics and electronic
noise, have been neglected.}
\end{figure}

Fig. \ref{fig:energy-loss} displays the simulated signal in the AT
as a function of incident beam energy. It shows the sum of the calculated
signals for all main-cell PMTs ($E_{\Sigma}$), for charged particles
produced by the four breakup modes of $^{4}\mathrm{He}$. Smearing
effects have been omitted to show the intrinsic signal more clearly.

In the MAX IV Laboratory experiment, tagged photons produced $^{4}\mathrm{He}$
photodisintegration events from breakup threshold up to 67~MeV. Neutrons
have a very small chance of interacting in the He gas and therefore
the distribution produced by $\gamma+{}^{4}\mathrm{He}\rightarrow n+^{3}\mathrm{He}$
is the simplest to interpret. There is one interacting ion, $^{3}\mathrm{He}$,
which stops more readily in the gas than \emph{$p$, $^{2}\mathrm{H}$
or $^{3}\mathrm{H}$} ions and therefore provides a means of calibrating
the AT energy response. The spread in $^{3}\mathrm{He}$ energy at
a given incident photon energy (Fig. \ref{fig:energy-loss}) results
from the spread in angle of the produced $^{3}\mathrm{He}$. The distribution
produced by $\gamma+{}^{4}\mathrm{He}\rightarrow p+^{3}\mathrm{H}$
is more complicated, with the ``cusp'' at $E_{\gamma}\sim25$~MeV
produced when protons cease to stop in the target gas, being the most
prominent feature. Three and four-body photodisintegration channels
produce relatively featureless distributions.

The AT is not able to separate the various photodisintegration processes
cleanly and the measured response is a convolution of the distributions
of Fig. \ref{fig:energy-loss}. MC calculations are compared with
the measured response in Sec. \ref{sec:MAX measurements}.

\subsubsection{\label{sub:AT-Detection-Efficiency}AT Detection Efficiency}

\begin{figure}
\begin{center}\includegraphics[width=1\columnwidth]{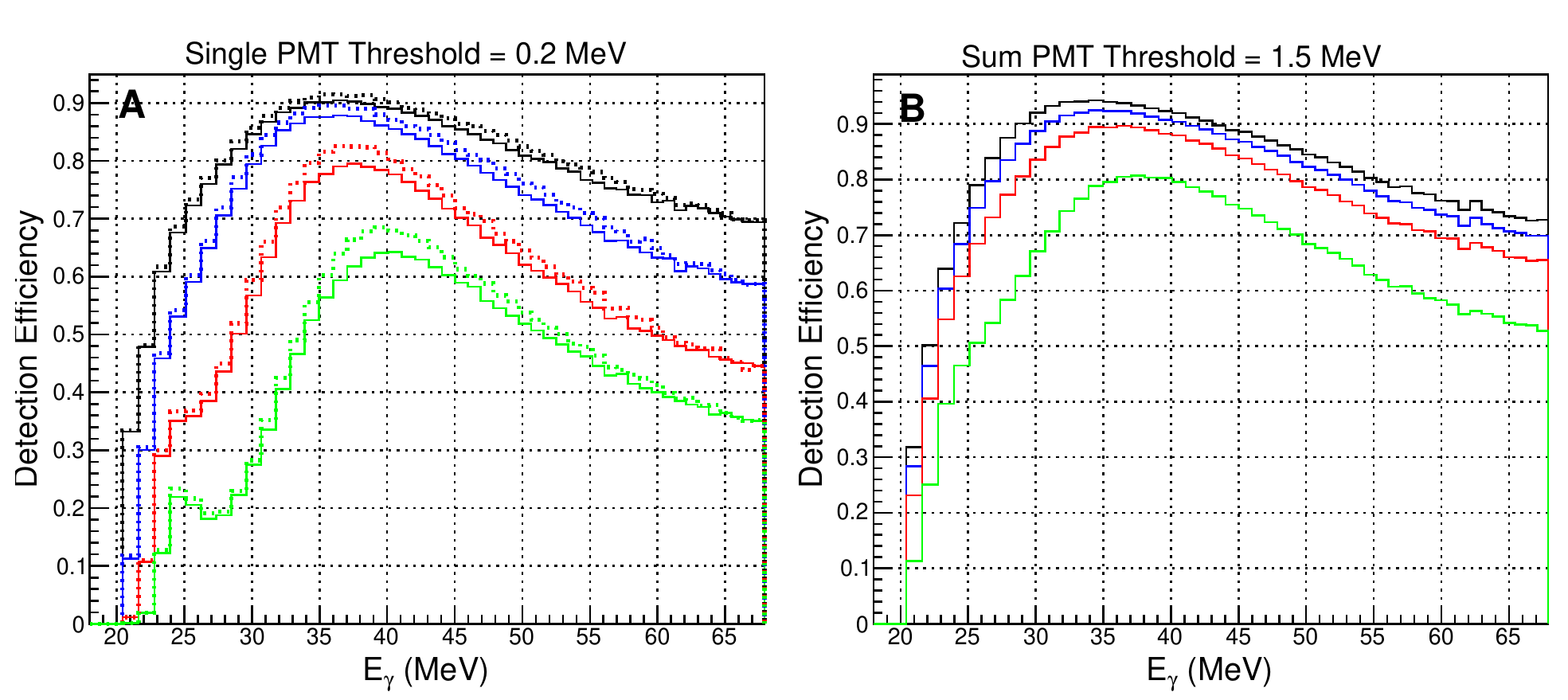}\end{center}

\protect\caption{\label{fig:AT-eff}A: MC detection efficiency as a function of incident
photon energy $E_{\gamma}$ at a fixed $T_{s}=0.2$~MeV and variable
$T_{\Sigma}$; black: 1~MeV; blue: 2~MeV; red: 3~MeV; green: 4~MeV.
Full-lines: isotropic two-body angular distributions, dotted-lines:
$\sin^{2}\theta$ two-body angular distributions. B: fixed $T_{\Sigma}=1.5$~MeV
and variable $T_{s}$; black: 0.0~MeV; blue: 0.1~MeV; red: 0.2~MeV;
green: 0.4~MeV. }
\end{figure}

The MC simulation was used to calculate the detection efficiency of
the AT for the range of photon energies employed in the MAX IV Laboratory
experiment. The energy deposited in the target is significantly different
for the different breakup modes of $^{4}\mathrm{He}$ (Fig.\ref{fig:energy-loss})
and close to threshold there is a strong dependence on photon energy.
Thus it is important that the reaction event generator models the
differential cross sections realistically. Here the employed partial
cross sections were obtained from Ref. \citep{Quaglioni}. Weighting
for the $\sim1/E_{\gamma}$ bremsstrahlung energy dependence has been
included. Pulse height resolution effects, due to photoelectron statistics
(5 photoelectrons per MeV in a single PMT (Sec.\ref{sub:Optical-Photon-Transport}))
and electronic noise (constant $\sigma=0.15$~MeV), were folded into
the simulation. Detection efficiency is just the fraction of reaction
events which produce a summed active target pulse height exceeding
a given detection threshold. MC generated data have been analysed
using the same event-selection conditions (Sec. \ref{sub:MAX analysis})
as the real data. Calculations (Fig.\ref{fig:AT-eff}) were made for
summed pulse height detection thresholds ($T_{\Sigma})$ in the range
0.5 -- 4~MeV (the hardware $T_{\Sigma}$ threshold was $\sim0.4$~MeV),
with the detection threshold for any single-PMT pulse height ($T_{s})$
fixed at values 0.0, 0.1, 0.2 and 0.4~MeV. One PMT only is required
to exceed $T_{s}$ for an accepted event and event time was not considered.
Non-constant bin widths in $E_{\gamma}$ follow the widths spanned
by each focal-plane detector of the MAX IV photon tagger. The bump
in efficiency observed at $E_{\gamma}\sim25$~MeV for thresholds
set above 3~MeV is related to the cusp in deposited proton energy
from the reaction $\gamma+^{4}\mathrm{He}\rightarrow p+^{3}\mathrm{H}$
(Fig.\ref{fig:energy-loss}). The sensitivity of the calculated efficiency
to the employed angular distributions for the two-body breakup channels
can be seen in Fig. \ref{fig:AT-eff}A. A $\sin^{2}\theta_{cm}$ distribution
(expected for incident dipole radiation) produces slightly higher
efficiency than an isotropic distribution, since particles produced
roughly perpendicular to the beam direction will on average see a
larger thickness of He gas.

\subsubsection{\label{sub:BeWindowSim}Simulation of Background Events}

MC calculations have been made to estimate the effect of photo reactions
on the Be windows which hold the gas pressure and the Al-Mylar windows
which isolate each cell of the AT optically. The Be has a total thickness
of 0.18~$\mathrm{g/cm^{2}}$, which is almost twice the total thickness
of the main AT gas cells (0.103~$\mathrm{g/cm^{2}})$. The Al-Mylar
(each window 0.5~$\mu$m Al evaporated on 5~$\mu$m $\mathrm{C_{10}H_{8}O_{4}}$)
has a total thickness of around 0.0035~$\mathrm{g/cm^{2}}$, about
3\% of the thickness of He gas. A detailed analysis of partial cross
sections for the nuclei of interest will be made when the $^{4}\mathrm{He}$
total photoabsorption cross section is evaluated. Here a rough estimate
of the relative integrated effect of photoreactions on the various
nuclei is given in Table \ref{tab:AT-materials}. $N_{nuclei}$ is
the number of nuclei per $cm^{2}$ relative to He, $\Sigma_{tot}=\varint\sigma_{tot}(E_{\gamma})dE_{\gamma}$
is the total photoabsorption cross section integrated over the range
$10<E_{\gamma}<100$~MeV \citep{Quaglioni,Ahrens} and $\mathcal{F}$
is the product $N_{nuclei}.\Sigma_{tot}$, parametrising the relative
numbers of photonuclear reactions produced in the in-beam materials.
The integrated effect of the 1000~ppm $\mathrm{N}_{2}$ gas admixture
is small relative to Al-Mylar.

\begin{table}
\begin{center}%
\begin{tabular}{|c|c|c|c|}
\hline 
Element & $N_{nuclei}$ & $\Sigma_{tot}$ (mb.MeV) & $\mathcal{F}$\tabularnewline
\hline 
\hline 
$_{2}^{4}\mathrm{He}$ & 1.0 & 105 \citep{Quaglioni} & 105\tabularnewline
\hline 
$_{4}^{9}\mathrm{Be}$ & 0.78 & 173 \citep{Ahrens} & 135\tabularnewline
\hline 
$_{6}^{12}\mathrm{C}$ & $7.1\times10^{-3}$ & 291 \citep{Ahrens} & 2.07\tabularnewline
\hline 
$_{7}^{14}\mathrm{N}$ & $1.0\times10^{-3}$ & 361 & 0.36\tabularnewline
\hline 
$_{8}^{16}\mathrm{O}$ & $2.8\times10^{-3}$ & 432 \citep{Ahrens} & 1.21\tabularnewline
\hline 
$_{13}^{27}\mathrm{Al}$ & $0.9\times10^{-3}$ & 739 \citep{Ahrens} & 0.67\tabularnewline
\hline 
\end{tabular}\end{center}

\protect\caption{\label{tab:AT-materials}Estimate of the integrated effect from photo
reactions on AT materials in the path of the photon beam. The effect
of H in the Mylar is negligible. Values of $\Sigma_{tot}$ were obtained
from Ref. \citep{Quaglioni,Ahrens}, apart from $^{14}\mathrm{N}$
where the value has been taken as the mean of the $^{12}\mathrm{C}$
and $^{16}\mathrm{O}$ values.}
\end{table}

\begin{figure}
\includegraphics[width=1\columnwidth]{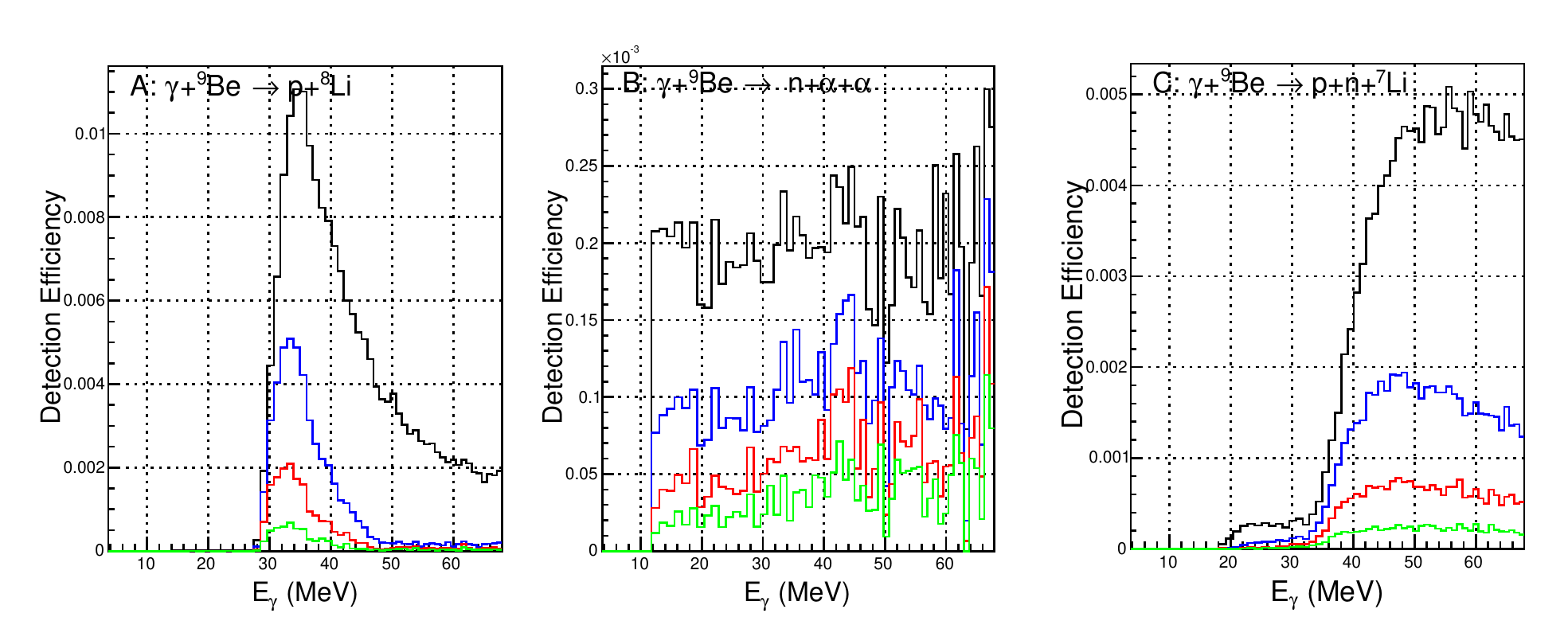}

\protect\caption{\label{fig:AT-Be-Windows}The AT detection efficiency as a function
of incident photon energy $E_{\gamma}$ for particles generated in
the two Be windows, at fixed $T_{s}=0.0$~MeV and variable $T_{\Sigma}$;
black: 1~MeV; blue: 2~MeV; red: 3~MeV; green: 4~MeV. }
\end{figure}

Window-isolation cells reduce the effect of photonuclear interactions
in the Be and the MC model has been used to estimate the detection
efficiency in the main AT cells for reaction products from $\gamma+^{9}\mathrm{Be}\rightarrow p+^{8}\mathrm{Li}$,
$\gamma+^{9}\mathrm{Be}\rightarrow p+n+^{7}\mathrm{Li}$ and $\gamma+^{9}\mathrm{Be}\rightarrow n+^{4}\mathrm{He}+^{4}\mathrm{He}$
photodisintegration channels. Calculations follow the method of Sec.
\ref{sub:AT-Detection-Efficiency} using angular distributions sampled
from kinematic phase space. However the reaction cross sections were
assumed to be independent of photon energy. Calculations were made
for both upstream and downstream windows and the combined results
are displayed in Fig.\ref{fig:AT-Be-Windows}. At $T_{\Sigma}=1$~MeV
the maximum detection efficiency for a $\gamma+^{9}\mathrm{Be}\rightarrow p+^{8}\mathrm{Li}$
event is slightly over 1\%. With low detection thresholds, the main
cell closest to the upstream window has a factor 5-6 more events than
its next neighbour, but this factor reduces as thresholds are raised.
The effect of Be on the measured reaction yield is assessed in Sec.
\ref{sub:AT-response} by comparison of yields obtained from the two
outer and two inner main AT cells.

\begin{figure}
\includegraphics[width=1\columnwidth]{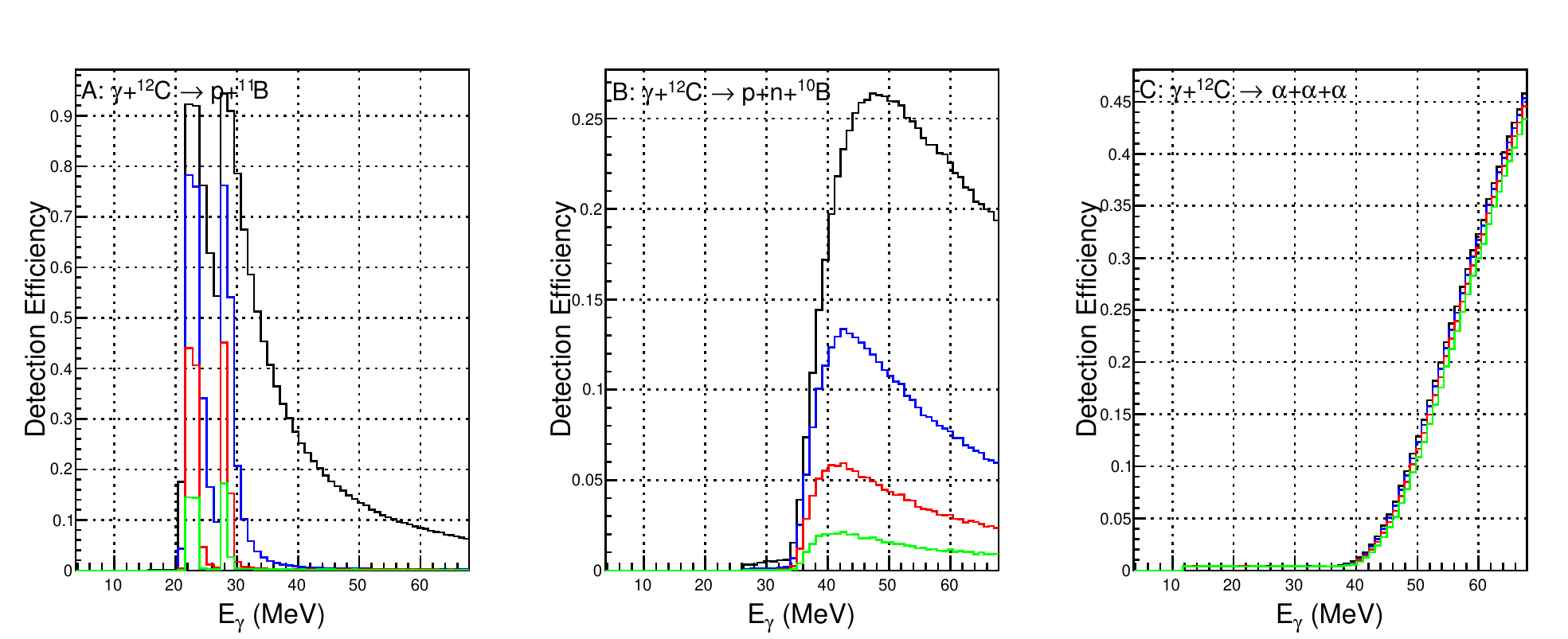}

\protect\caption{\label{fig:AT-Al-Mylar}The AT detection efficiency as a function
of incident photon energy $E_{\gamma}$ for particles generated in
an Al-Mylar window, at fixed $T_{s}=0.0$~MeV and variable $T_{\Sigma}$;
black: 1~MeV; blue: 2~MeV; red: 3~MeV; green: 4~MeV. }
\end{figure}

The Al-Mylar windows are thin compared to the Be (Table \ref{tab:AT-materials}),
but on the other hand they are directly adjacent to the main-cell
gas and so photo reaction particles are much more readily detected.
Carbon has the highest $\mathcal{F}$ factor of the Al-Mylar elements
and Fig. \ref{fig:AT-Al-Mylar} displays MC calculations for $\gamma+^{12}\mathrm{C}\rightarrow p+^{11}\mathrm{B}$,
$\gamma+^{12}\mathrm{C}\rightarrow p+n+^{10}\mathrm{B}$ and $\gamma+^{12}\mathrm{C}\rightarrow^{4}\mathrm{He}+^{4}\mathrm{He}+^{4}\mathrm{He}$.
The method follows that employed for Be. Single $p$ knock out is
most important at giant-dipole-resonance energies, with $pn$ knock
out becoming more important (both in terms of efficiency and cross
section) at higher energies. The 3$\alpha$ cross section is likely
to be very small \citep{3-alpha} in the region where its efficiency
becomes significant. Measured yields (Sec.\ref{sub:AT-response})
suggest that some non-helium background is detected (especially at
low threshold settings) and more detailed MC event generators, which
model partial reaction channel differential cross sections as realistically
as possible, will be necessary to obtain a more quantitative estimate
of this background.

\section{\label{sec:MAX measurements}Measurements of $^{4}$He Photodisintegration
at MAX IV Laboratory.}

\textcolor{black}{The electron accelerator at MAX IV Laboratory \citep{MAX-lab},
operating in pulse-stretcher mode for nuclear physics experiments,
delivers a $\sim50\%$ duty-factor beam up to an energy of $\sim200$~MeV
and a current of $\sim30$ nA. Bremsstrahlung, produced on a thin
Al foil, is tagged by detecting the post-bremsstrahlung electron,
either in the broad-band main tagging (MT) spectrometer \citep{MT}
or in the end-point tagging spectrometer. The present experiment used
an electron beam energy of }164.7~MeV, with the MT set for a central
momentum of 94.8~MeV/c, producing a tagged-photon energy range \textcolor{black}{of
$E_{\gamma}=4.4-67.5$~MeV. The focal plane detector of the MT was
segmented into 62 channels, giving an average channel width of $\sim1$~MeV.}

\begin{figure}
\includegraphics[width=1\columnwidth]{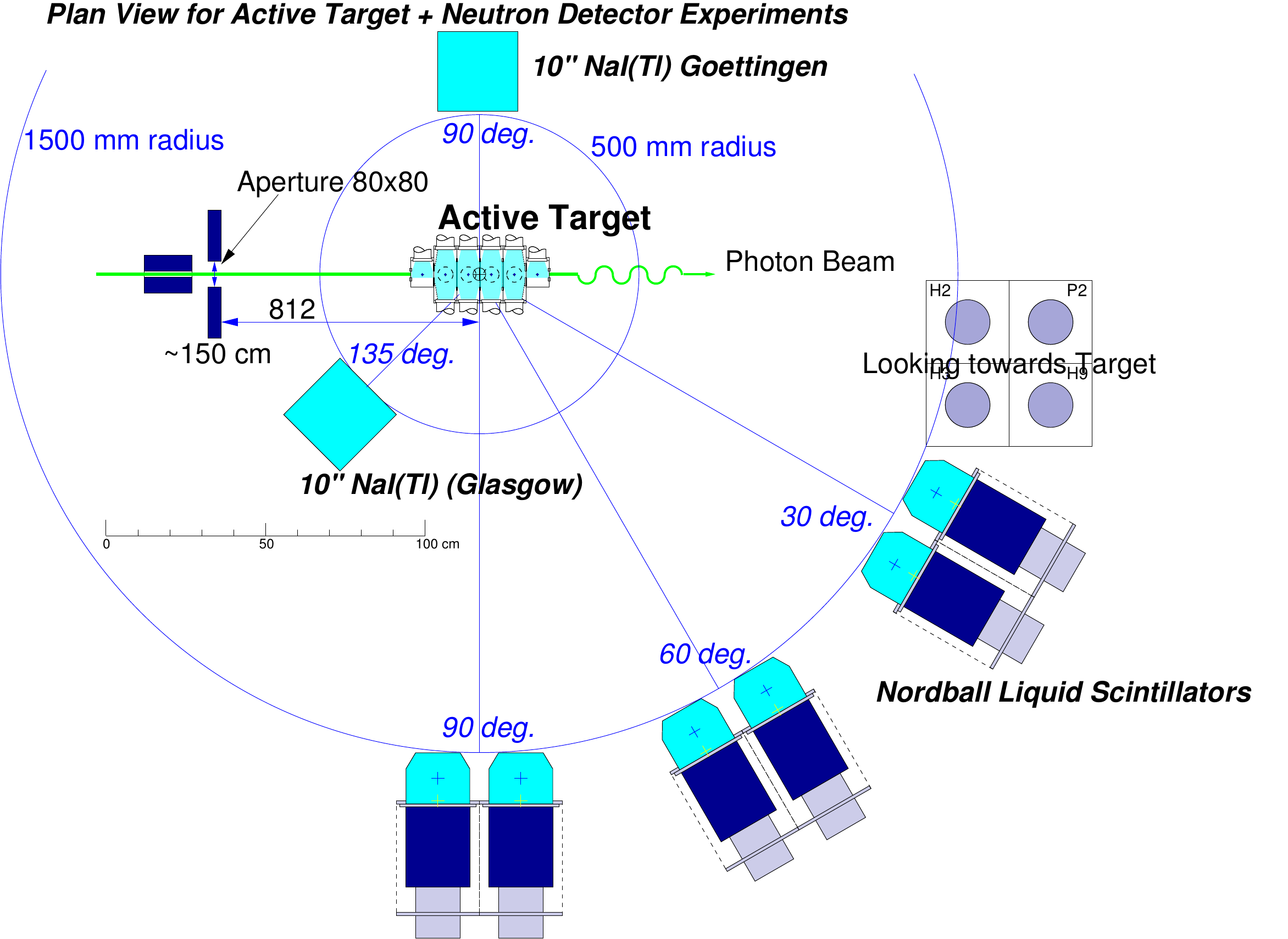}

\protect\caption{\label{fig:AT-plan-view}Plan of the AT experiment at MAX IV Laboratory.}
\end{figure}

\textcolor{black}{The AT was placed directly in the photon beam (Fig.
\ref{fig:AT-plan-view}) which was collimated to produce a spot of
11~mm diameter on the entrance window. }As the AT is relatively insensitive
to the electrons produced in pair-production or Compton-scattering
processes and has a sharp scintillation pulse (duration $\sim20$~ns)
it ran comfortably up to the maximum available photon beam intensity.
A time resolution of $\sim1$ ns was obtained (Sec.\ref{sub:MAX analysis}),
giving $\sim2:3$ signal-to-random ratio for coincidences with the
tagger focal-plane detectors, which at maximum intensity counted at
average rates in excess of 1~MHz. Measurements were made at maximum
intensity, to obtain reasonable numbers of coincidences between the
AT and external detectors, and also at a factor 10 lower intensity
for inclusive measurements of the $\mathrm{^{4}He}$ total photo absorption
cross section. 

Good timing performance makes the AT suitable as a ``start'' detector
for neutron time of flight (TOF) measurement. \textcolor{black}{The
TOF spectrometer, the ``Nordball'' array \citep{nordball_tof} of
liquid scintillators (Fig.\ref{fig:AT-plan-view}), was positioned
at angles $30,\:60,\:90^{\circ}$ and a flight path of $\sim1.5$~m
to measure coincident neutrons produced by $\gamma+{}^{4}\mathrm{He\rightarrow n+^{3}\mathrm{He}}$,
$\gamma+{}^{4}\mathrm{He\rightarrow\mathit{n+p+d}}$ and $\gamma+{}^{4}\mathrm{He\rightarrow\mathit{2n+2p}}$
reactions in the target. Neutron energy was measured by TOF \citep{jrma_tof},
with the AT providing the time reference, and pulse shape analysis
\citep{jrma_psd} was employed to distinguish interacting neutrons
from photons or electrons. With detection thresholds set to 100~keV
electron equivalent, Nordball can detect neutrons of energy above
$\sim1$~MeV \citep{Reiter,Reiter1}.}

Two 10'' NaI(Tl) counters were positioned at angles $90,\:135^{\circ}$
to detect energetic photons from nuclear Compton scattering events.
The coincident recoiling $^{4}\mathrm{He}$ in the AT, together with
the tagged photon energy would in principle allow full reconstruction
of the Compton scattering kinematics, enabling efficient rejection
of background processes which otherwise contaminate the very weak
Compton signal. Analysis of this data is not presented here.

\subsection{\label{sub:MAX analysis}Collection and Analysis of Active Target
Data}

Pulses from the 4 PMTs of each main AT cell were summed and fed to
discriminators, to produce the data readout trigger. This also gave
the start and gate signals for all time- and charge-to-digital converters
(TDC, QDC). Photon beams produce a high flux of electron background
from atomic interactions of photons with the target gas, entrance/exit
windows and air. However the average $dE/dx$ of electrons in 2~MPa
He is $\sim6$~keV/cm so that the target is rather insensitive to
this background.

Background was observed in the AT, which was not due to He scintillations,
since it was present even with the AT evacuated. Generally it had
larger amplitude than PMT dark-current noise and has a number of possible
sources:
\begin{itemize}
\item electron or positron interactions with the dynodes of the PMTs.
\item electron or positron interactions in the quartz AT windows.
\item electron or positron interactions with the glass PMT cathode windows.
\item neutron interactions with the Boron in the PMT glass.
\end{itemize}
Empty cell measurements show that non-scintillation processes generally
produce a significant pulse height in only one of the four PMTs attached
to an AT cell and can be suppressed by demanding that >1 PMT from
a cell has fired. All 18 individual PMT signals are attached to discriminators
so that the PMT multiplicity can optionally be incorporated in the
trigger. Liquid scintillator and NaI(Tl) signals were not included
in the trigger, but their pulse amplitudes and hit times were recorded. 

The MAX I ring, which provided stretched beam for the tagged-photon
facility, operated with a 10~Hz injection rate from a pulsed LINAC.
During the first ms of the 100~ms long extraction cycle there is
a sharp spike in the beam intensity and thus the electronics were
inhibited during this period.

\begin{figure}
\includegraphics[bb=0bp 230bp 567bp 518bp,clip,width=1\columnwidth]{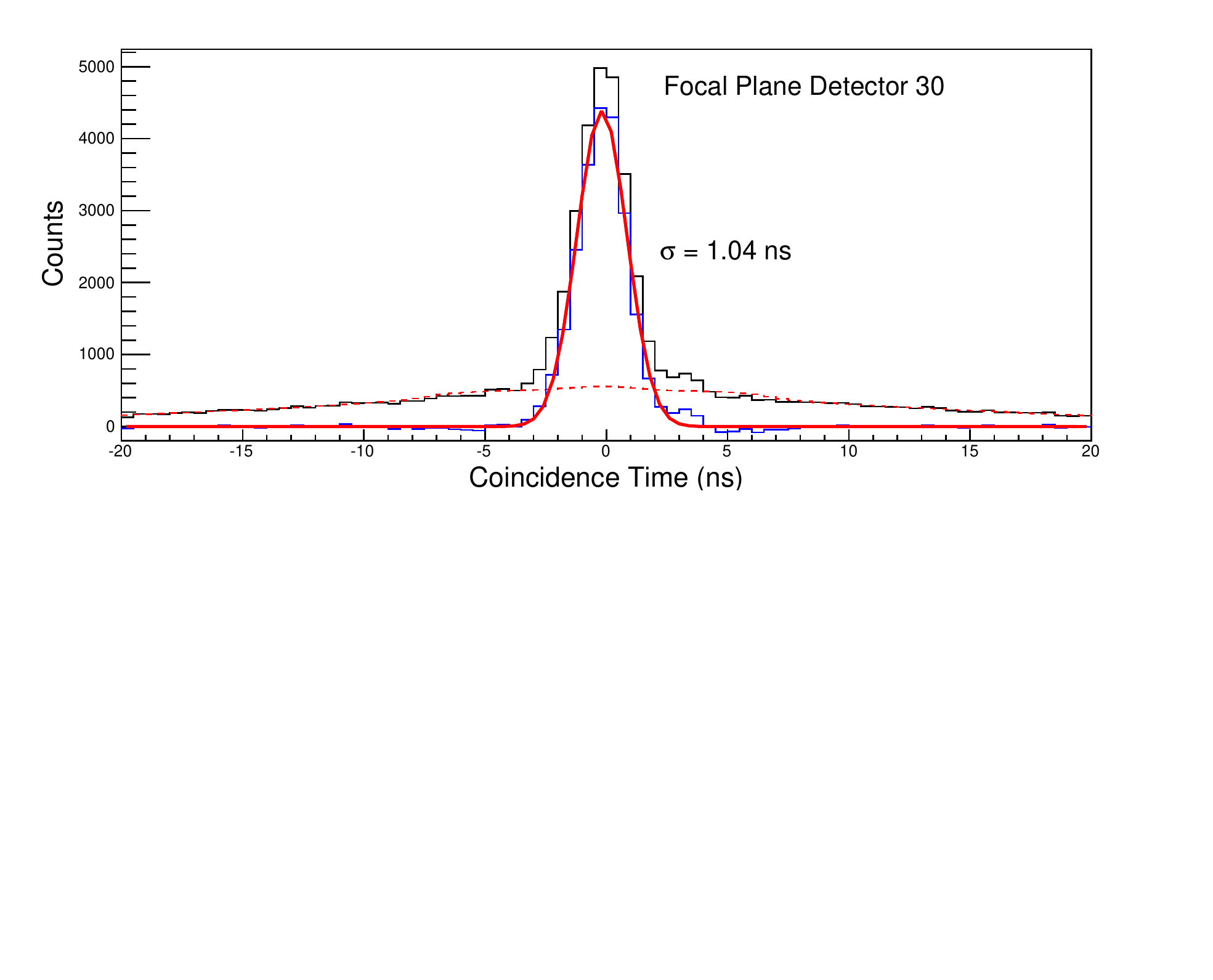}

\protect\caption{\label{fig:Time-FPD}Time spectrum of coincidences between the AT
and element 30 of the FP hodoscope made during a low-intensity run.
The difference of the ``raw'' spectrum (black) and the random background
estimate (red dashed) produces the background-subtracted spectrum
(blue) which is fitted with a Gaussian (full red line).}
\end{figure}

\begin{figure}
\includegraphics[clip,width=1\columnwidth]{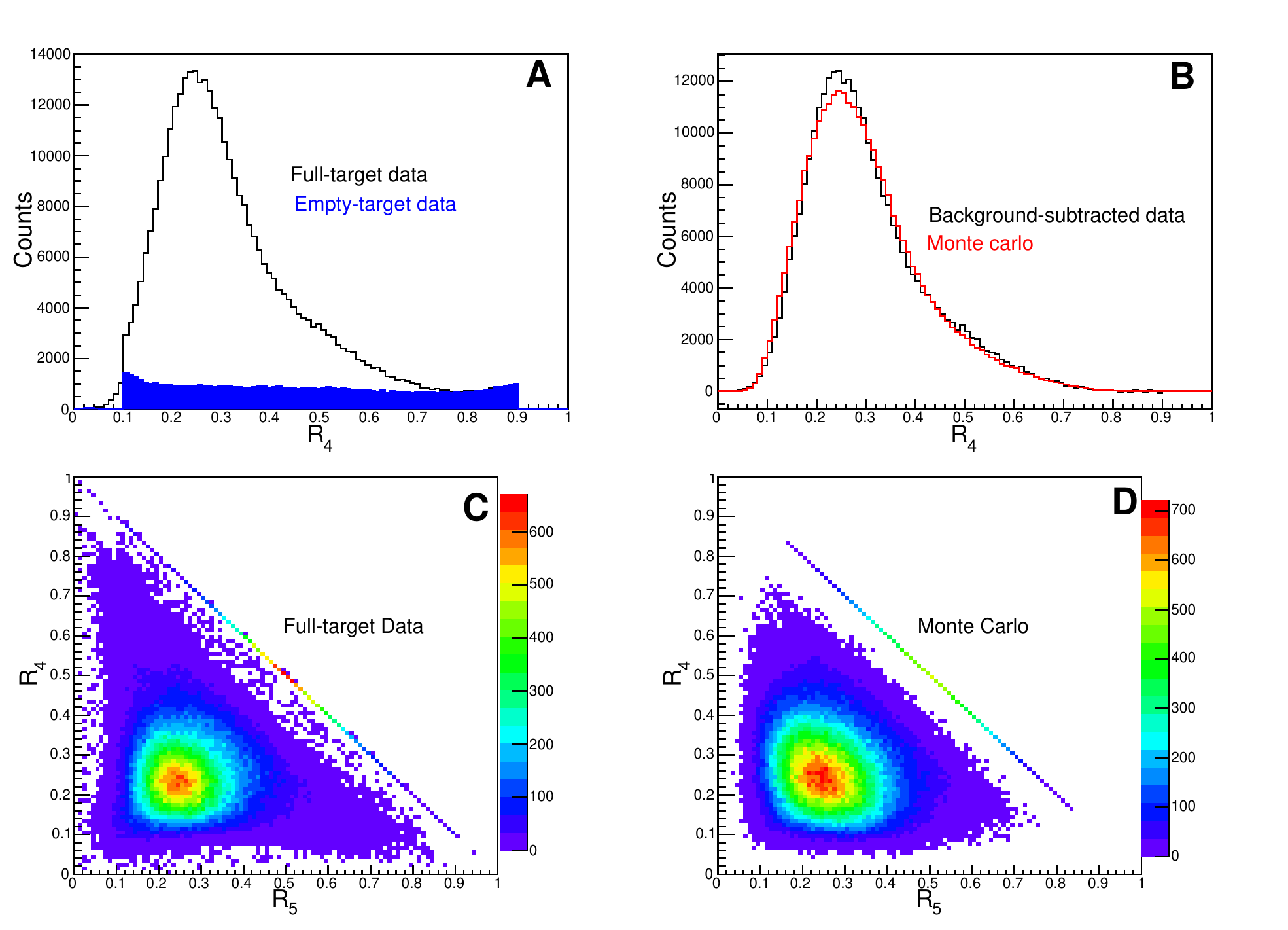}

\protect\caption{\label{fig:Ratio-plots}Ratio plots for signals from PMT 4 ($R_{4}$)
and 5 ($R_{5}$) (Fig. \ref{fig:Active-Target-geometry}). Plot A
shows an empty-target background subtraction; B compares a background-subtracted
measurement with a MC prediction; C shows a measurement without background
subtraction; D shows a Monte Carlo prediction. The diagonal lines
in 2D plots C,D correspond to events where only PMT 4 and 5 produced
a significant signal.}
\end{figure}

The following parameters were recorded for off-line reaction reconstruction.
\begin{enumerate}
\item Hit times (relative to the active target) of the 62 plastic scintillators
which make the focal-plane (FP) detector of the tagger were recorded
in multi-hit TDCs. A coincident hit (Fig. \ref{fig:Time-FPD}) in
a particular FP counter corresponds to a particular photon energy
with a channel width of $\sim1$~MeV. The random coincidence background
has been estimated using the technique of Ref.\citep{Morhac} implemented
through the ``TSpectrum'' class of the ROOT analysis library \citep{root}.
After subtraction of the background, the integral from -10 to +10~ns
of the resulting coincidence peak gives the AT yield for a particular
energy bin. The peak in Fig. \ref{fig:Time-FPD} has been fitted with
a Gaussian producing a width $\sigma=1.04$~ns. Fitted widths for
the other FP detectors fall in the range $1.0<\sigma<1.5$~ns.
\item Pulse charge and relative time from all 18 PMTs on the active target
were recorded in QDCs and TDCs. The scintillations should produce
similar signals in each PMT, while electron interactions in optical
windows (Cherenkov light) or the PMT electrodes will produce a disproportionately
large signal in a single PMT. Thus the balance of charge can be used
to select scintillations. The parameter $R_{i}=Q_{i}/\sum Q_{j}$,
where $Q_{i}$ is the charge produced by a single photomultiplier
and $\sum Q_{j}$ is the sum of the four PMTs attached to a particular
AT cell, should produce a distribution centred at $R_{i}\sim0.25$.
Fig.~\ref{fig:Ratio-plots} compares the measured response with that
calculated by the AT Geant-4 model, assuming only scintillation processes
in the He gas. The scintillation signal falls within the range $0.1<R_{i}<0.7$
and the real data has been filtered so that at least one PMT combination
is within these limits of $R_{i}$, to suppress spurious, single-PMT
hits. 
\item Time and amplitude from each of the 12 liquid-scintillator elements
of the Nordball array were recorded in QDCs and TDCs. The pulse-shape
signal was also recorded in a voltage-to-digital converter (VDC).
The neutron TOF signal determines the momentum and signal quality
may be enhanced by an off-line cut on the pulse-shape information. 
\item Time and amplitude from the two NaI(Tl) counters were recorded in
TDCs and QDCs.
\end{enumerate}

\begin{comment}
\begin{figure}
\begin{centering}
\includegraphics[width=1\columnwidth,keepaspectratio,bb = 0 0 200 100, draft, type=eps]{Plots/ActiveTarget.eps}
\par\end{centering}

\protect\caption{\label{cap:Offline-selection}A,B selection of the target scintillation
signal, C selection of neutrons in Nordball, D tagged-photon coincidences.}
\end{figure}
\end{comment}

\subsection{\label{sub:AT-response}The AT Pulse Height Response}

\begin{figure}
\includegraphics[width=1\columnwidth]{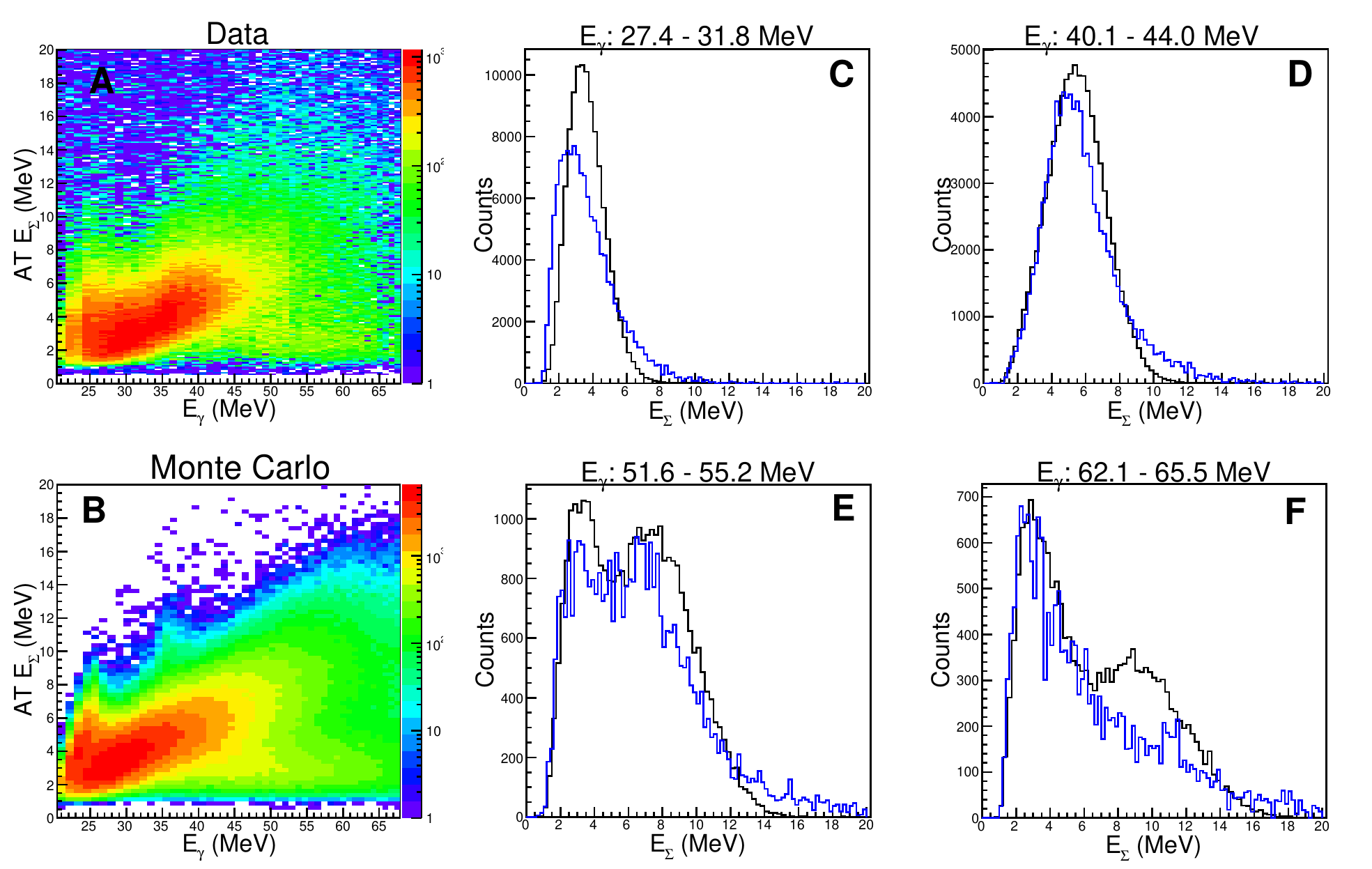}

\protect\caption{\label{fig:AT-PH-Eg}A: the measured AT $E_{\Sigma}$ as a function
of $E_{\gamma}$. B: the equivalent MC prediction. 1D projections
of the 2D distributions (vertical slices in a given range of $E_{\gamma}$)
are given in C, D, E, F, with data in blue and MC in black.}
\end{figure}

The measured AT pulse height response, as a function of tagged photon
energy, is compared with the MC calculation in Fig. \ref{fig:AT-PH-Eg}.
Random coincidence contributions to the AT pulse height distributions
(see Fig. \ref{fig:Time-FPD}) have been subtracted. There is some
variability in the efficiency of the tagger channels, which produces
discontinuities in the experimental data. The main features of the
measured distribution are reproduced by the simulation, which is binned
in $\mathrm{E}_{\gamma}$ identically to the measurement. The band
of $E_{\Sigma}$ produced by recoiling $^{3}\mathrm{He}$ and $^{3}\mathrm{H}$
ions, which increases with $\mathrm{E}_{\gamma}$, is clearly visible.
Similarly the ``cusp'' at $\mathrm{E}_{\gamma}\sim24$~MeV, produced
as protons cease to stop in the He gas is seen, although the second
weaker cusp at $E_{\gamma}\sim35$~MeV is less evident in the experimental
data where the statistical fluctuations are greater. In the 1D projections
showing $E_{\Sigma}$ distributions for specific bites of $E_{\gamma}$,
the MC has been normalised to the integral of the data distribution.
There are some differences in the shapes which will have the largest
effect on detection efficiency (Fig. \ref{fig:AT-PH-Eg}C) at low
$E_{\Sigma}$, close to where thresholds are applied. At higher photon
energies (Fig. \ref{fig:AT-PH-Eg}F) the predicted bump at $E_{\Sigma}\sim10$~MeV
is not visible in the data, suggesting that the $n+{}^{3}\mathrm{He}$
cross section input to the MC is too high at $E_{\gamma}\apprge60$~MeV.

\begin{figure}
\includegraphics[width=1\columnwidth]{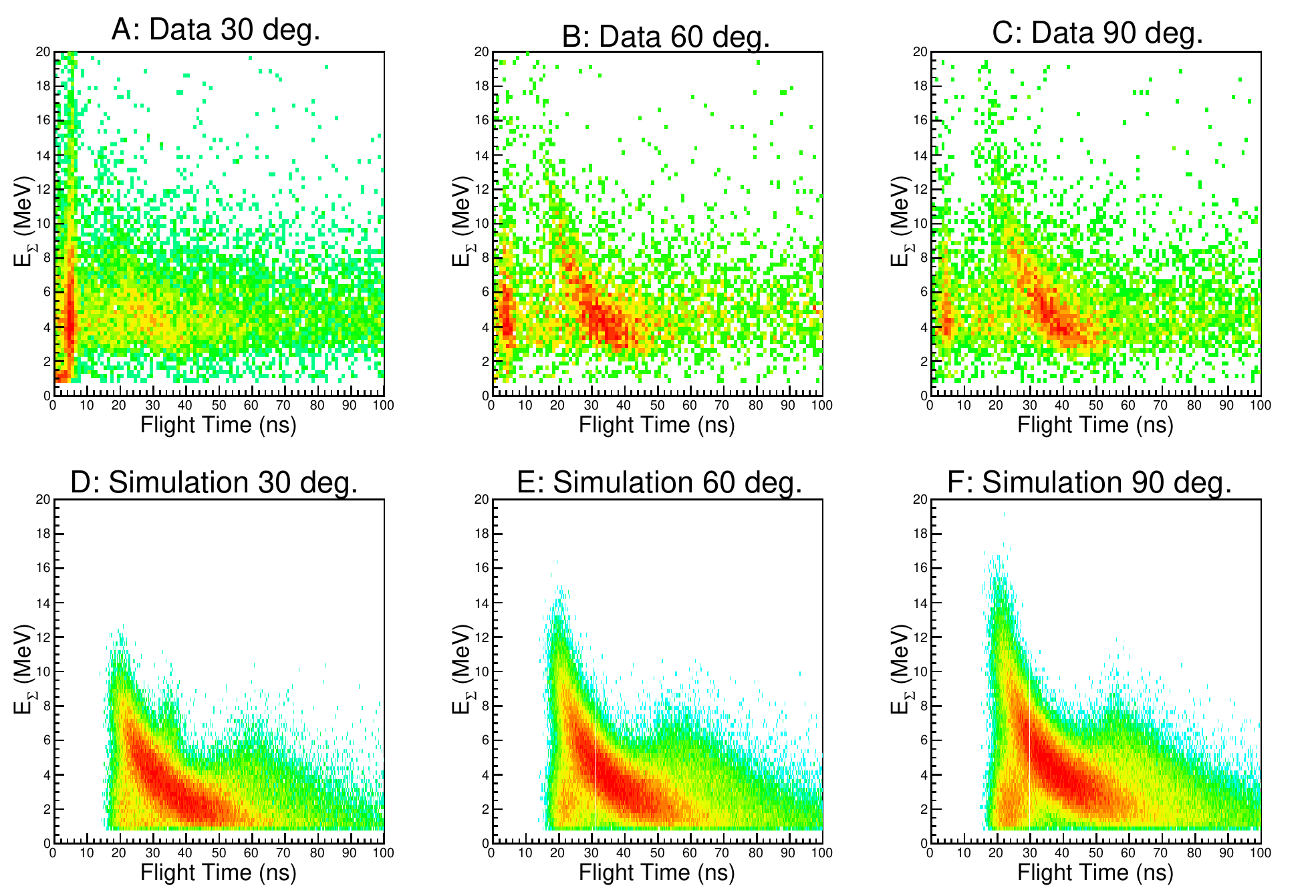}

\protect\caption{\label{fig:Comp-TOF}Comparison of measured (A,B,C) AT summed energy,
$E_{\Sigma}$, as a function of coincident neutron TOF, with that
calculated by the Geant-4 model (D,E,F). The neutron angles are $30\pm5^{\circ}$
(A,D), $60\pm5^{\circ}$ (B,E), $90\pm5^{\circ}$ (C,F), }

\end{figure}

Fig. \ref{fig:Comp-TOF} displays the AT pulse height ($E_{\Sigma})$
correlation with NordBall TOF. The prominent curved band, corresponding
to the $n+\mathrm{^{3}He}$ final state, is clearly seen in both measurement
and MC. These are reasonably consistent, showing the increase in maximum
$^{3}\mathrm{He}$ pulse height at TOF $\sim20$~ns, as the neutron
angle increases from $30-90^{\circ}$. The MC calculation does not
include scattered relativistic photons or electrons, which show as
a vertical band in the experimental data at flight times $\sim5$~ns. 

\subsection{Systematic Corrections to the Reaction Yield}

The inclusive $\gamma+^{4}\mathrm{He\rightarrow X}$ yield as a function
of $\mathrm{E_{\gamma}}$ was determined by integrating the random-subtracted
coincidence peaks in the 62 FP time spectra. Data filtering (see Fig.\ref{fig:Ratio-plots})
to select He scintillation events and subtraction of non-scintillation
empty-cell background was applied. This yield was then normalised
to units of $\mu$b using the relation:

\begin{equation}
Y=\frac{N_{tdc}}{\varepsilon_{d}\,\varepsilon_{tagg}\, N_{e}\, N_{T}}\label{eq:Yx}
\end{equation}

\begin{figure}
\includegraphics[width=1\columnwidth]{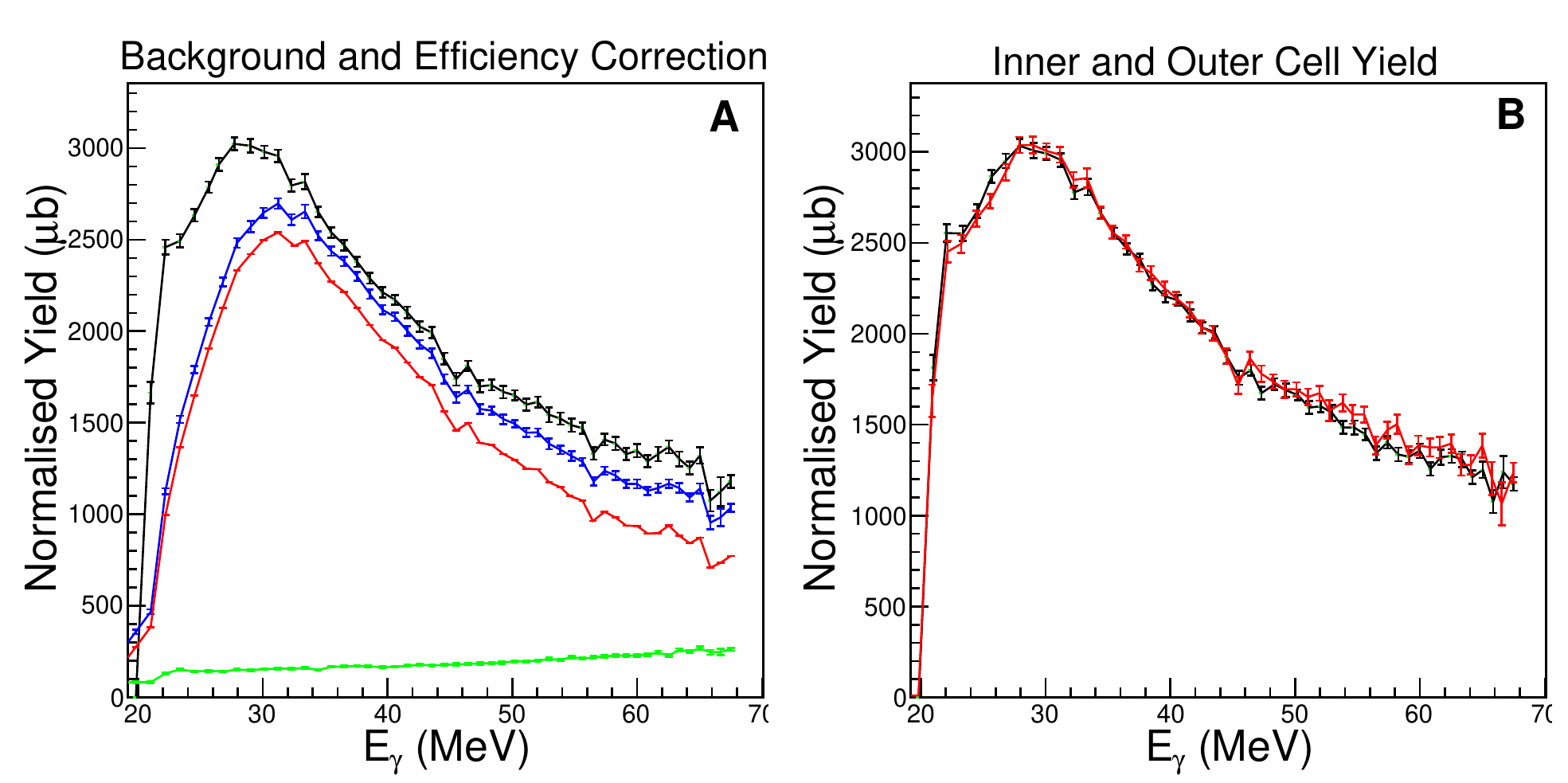}

\protect\caption{\label{fig:AbsYield}AT normalised yield $\mathit{Y}$ at a fixed
$T_{s}=0.2$~MeV, $T_{\Sigma}=1.5$~MeV. A) blue: raw yield; green:
non-scintillation background yield; red: background-subtracted yield;
black efficiency corrected yield. B) corrected yields, black: inner
two AT main cells; red outer two AT main cells, thresholds as A).}
\end{figure}

The parameters of Eq.\ref{eq:Yx} are explained as follows:
\begin{itemize}
\item $N_{tdc}$ is the yield obtained from integration of the coincidence
peak of a  tagger TDC (Fig.\ref{fig:Time-FPD}). Since the inclusive-yield
runs were made at relatively low intensity using multi-hit TDCs, dead
time corrections to the yield were not necessary. 
\item $\varepsilon_{d}$ is the AT detection efficiency, obtained from the
MC simulation assuming phase-space angular distributions for $^{4}\mathrm{He}$
photoreactions, as described in Sec.\ref{sub:AT-Detection-Efficiency}. 
\item $\varepsilon_{tagg}$ is the tagging efficiency, the probability of
a bremsstrahlung photon passing through the collimator, given a hit
in the FP detector. The tagging efficiency was measured periodically
by inserting a 100\% efficient Pb-scintillating fibre detector in
the photon beam and counting coincidences at reduced beam intensity.
The tagging efficiency was determined from the mean of 12 measurements
taken periodically between the main runs and was found to vary systematically
with photon energy between 8 and 9\%.
\item $N_{e}$ is the number of electrons registered in the tagger, counted
by the scalers attached to particular tagger FP detectors. The number
of tagged photons incident on the target is $\varepsilon_{tagg}N_{e}$.
\item $N_{T}$ is the number of target nuclei per $cm^{2}$ at 2~MPa pressure.
Both AT pressure and temperature were monitored continuously and the
overall uncertainty due to gas pressure is estimated at $\sim5$\%.
Around $\sim2$\% will be possible when run-by-run pressure corrections
are made. Since the Be windows of the target do not ``bow'' significantly
under pressure, the length of gas in the target (288~mm total) could
be determined to better than 1~mm.
\end{itemize}
Fig.\ref{fig:AbsYield}A displays the effect on $Y$ of background
and efficiency corrections, where the error bars show the statistical
uncertainties. The size of these corrections depends on the applied
$T_{s}$ and $T_{\Sigma}$: respectively the single and summed PMT
energy thresholds. The background increases and the efficiency correction
decreases as the thresholds are lowered. Fig.\ref{fig:AbsYield}B
compares $Y$ from the two outer and two inner AT main cells. There
is no significant difference, consistent with an insignificant background
from the Be windows.

\begin{figure}
\includegraphics[width=1\columnwidth]{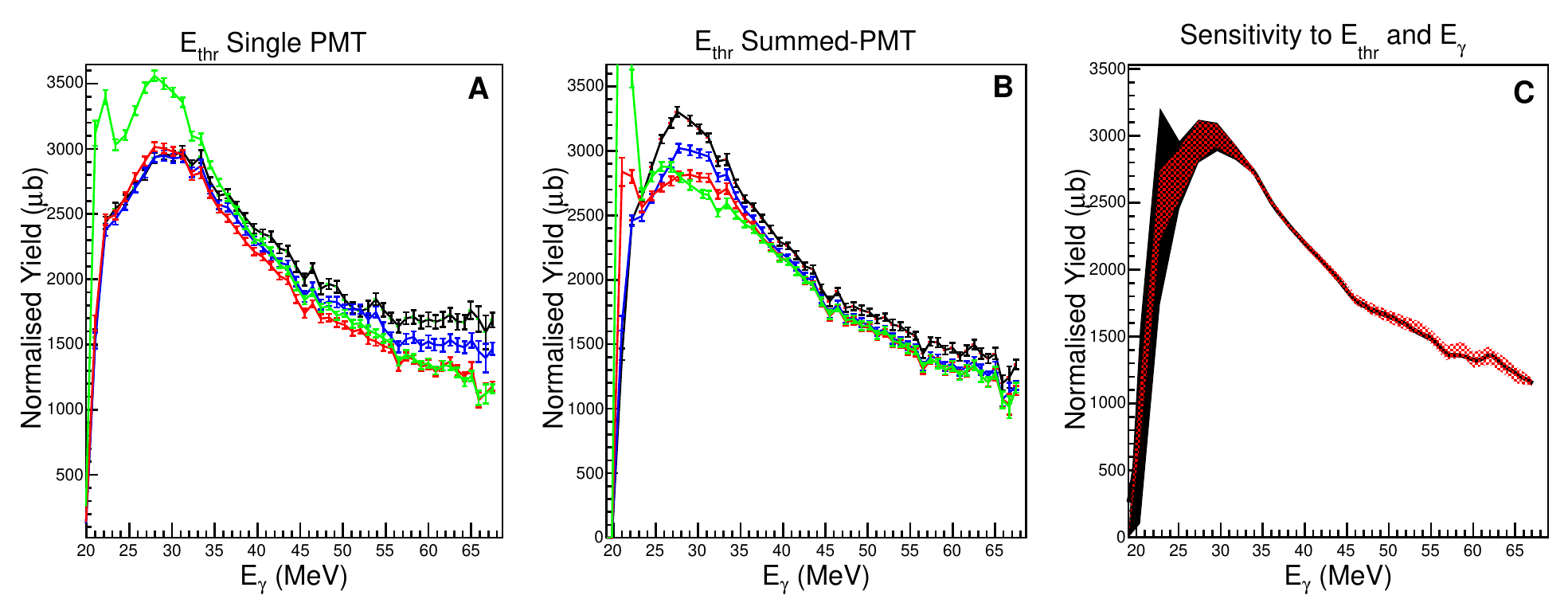}

\protect\caption{\label{fig:AbsYield1}AT normalised yield. A) at a fixed $T_{\Sigma}=1.5$~MeV,
black: $T_{s}=0.0$~MeV; blue: $T_{s}=0.1$~MeV; red: $T_{s}=0.2$~MeV;
green: $T_{s}=0.4$~MeV. B) at a fixed $T_{s}=0.2$~MeV, black:
$T_{\Sigma}=1.0$~MeV; blue: $T_{\Sigma}=1.5$~MeV; red: $T_{\Sigma}=2.0$~MeV;
green: $T_{\Sigma}=2.5$~MeV. C) Error bands showing sensitivity
to uncertainties in $E_{\gamma}$ (black filled) and $T_{\Sigma}$(red
shaded).}
\end{figure}

Fig.\ref{fig:AbsYield1} displays the effect on $Y$ of varying $T_{s}$
and $T_{\Sigma}$. In panel A $T_{\Sigma}$ is fixed at 1.5~MeV.
At $E_{\gamma}\lesssim40$~MeV, $Y$ is relatively stable with respect
to $T_{s}$, apart from $T_{s}=0.4$~MeV where the efficiency correction
appears to be too large. Above 40~MeV, $Y$ drops as $T_{s}$ increases
to 0.2~MeV, but stabilises between 0.2 and 0.4~MeV. With $T_{s}$
fixed at 0.2~MeV (Fig.\ref{fig:AbsYield1}B) $Y$ is quite stable
with respect to change in $T_{\Sigma}$ at $E_{\gamma}\gtrsim35$~MeV,
but there is significant variation at lower energies. At $E_{\gamma}<25$~MeV
the efficiency correction is apparently too large for $T_{\Sigma}\geq2.0$~MeV.
Panels A and B show statistical uncertainties, while C gives an estimate
of the sensitivity of the corrected yield to systematic effects at
$T_{s}=0.2$~MeV, $T_{\Sigma}=1.5$~MeV. The black error band shows
the effect of a systematic error of $\pm1$~MeV in the calibration
of $E_{\gamma}$, which is large below 30~MeV. The red error band
shows the effect of a systematic error of $\pm0.25$~MeV in the calibration
of $T_{\Sigma}$. Again this is largest at lower photon energies.

Although $Y$ is given in $\mu$b, it is not yet a cross section.
Overall $Y$ has greatest stability with respect to variation in threshold
at $T_{s}\sim0.2$~MeV and $T_{\Sigma}\sim1.5$~MeV, but it is apparent
that the MC calculation of detection efficiency requires improvement.
For $^{4}\mathrm{He}$ the sensitivity of the calculated efficiency
to the relative cross sections for the different breakup channels
needs to be examined more carefully: for example the lack of observed
$\gamma+^{4}\mathrm{He}\rightarrow n+^{3}\mathrm{He}$ signal compared
to MC at higher $E_{\gamma}$. Data for this and the three- and four-body
breakup channels are rather sparse at $E_{\gamma}\gtrsim40$~MeV.
More realistic modelling of the angular distributions may also be
necessary as the present calculations show a modest sensitivity in
this respect. The Be windows appear not to have a significant effect,
but contributions from the Al-Mylar windows remain to be evaluated
in detail. The integrated effect cannot be large (Sec.\ref{sub:BeWindowSim}),
but different (from $^{4}\mathrm{He}$) $E_{\gamma}$ dependence of
differential cross sections could produce significant localised effects,
especially close to $^{4}\mathrm{He}$ breakup threshold.

\section{\label{sec:Summary}Summary and Outlook}

A He gas scintillator has been developed for photonuclear reaction
experiments. The scintillator operates at a pressure of 2~MPa and
uses an admixture of 1000~ppm of $\mathrm{N}_{2}$ to shift the primary
scintillation from UV to visible wavelengths. When used as an active
target, this level of admixture keeps $\mathrm{N}_{2}$ contributions
to cross section measurements below the 1\% level in general. The
scintillation pulse has a rise time of $\sim5$~ns and fall time
of $\sim10$~ns, so that the detector can provide a precise time
reference and run at relatively high rates.

An experiment at the MAX IV Laboratory tagged photon facility has
shown that the He gas scintillator can be used as an active target
for photonuclear studies. Runs at full photon beam intensity demonstrated
the timing and rate capability of the AT. In addition they produced
sufficient coincident counting rate to correlate the AT signal with
neutral particles escaping and interacting in an external TOF spectrometer.
This detector system has already been used to measure the photon asymmetry,
$\Sigma,$ of $\gamma+^{4}\mathrm{He}\rightarrow n+^{3}\mathrm{He}$
\citep{Ganenko} with linearly-polarised incident photons.

The ability of the AT to provide an accurate measurement of the total
$^{4}\mathrm{He}$ photoabsorption cross section was assessed from
data taken at a factor $\sim10$ lower beam intensity, where rate-dependent
corrections to yields are very small, and an open trigger may be used
without overwhelming the data acquisition system. Even at reduced
intensity, excellent statistical uncertainties were obtained with
a few days of running. Similarly small systematic uncertainties can
be achieved for $E_{\gamma}\gtrsim35$~MeV, but as breakup threshold
($E_{\gamma}\sim20$~MeV) is approached the size of the detection
efficiency correction, and its uncertainty, increases. Work is in
progress to finalise the tagged-photon energy calibration and to extend
the MC event generators to give a more realistic description of photoreactions
on He and other AT materials. These will both have a bearing on the
final evaluation of the $^{4}\mathrm{He}$ total photoabsorption cross
section and its systematic uncertainty.

In the future it is planned to use a gas scintillator to measure $\gamma+^{3,4}\mathrm{He}\rightarrow\gamma+^{3,4}\mathrm{He}$
in conjunction with the Crystal Ball electromagnetic calorimeter at
Mainz \citep{mami}. The possibility to use Si photomultipliers inside
the pressure vessel to detect the scintillation is being investigated.

\subsection*{Acknowledgements}

The authors acknowledge the outstanding support of the MAX IV Laboratory
staff. We are also grateful for the support of the UK Science and
Technology Facilities Council (grant numbers 57071/1, 50727/1 and
ST/H003177/1), the Swedish Research Council, the Knut and Alice Wallenberg
Foundation and the Crafoord Foundation.

\end{document}